\documentclass[useAMS,usenatbib]{mnras}

\usepackage[dvipsnames]{xcolor}

\usepackage{graphicx}\usepackage{epsfig}\usepackage{epsf}
\usepackage{amsmath}\usepackage{amssymb}\usepackage{stfloats}
\usepackage{soul}
\usepackage{longtable}
\newcommand\D[1]{#1} 

\newcommand\reftwo[1]{#1} 

\usepackage{caption}
\usepackage{subcaption}

\title[The Superluminous Type IIn Supernova ASASSN-15ua]{The Superluminous Type IIn Supernova ASASSN-15ua: Part of a continuum in extreme precursor mass loss}

\author[Dickinson et al.]{Danielle Dickinson$^{1,2}$\thanks{E-mail:
    dickinsd@purdue.edu}, Nathan Smith$^{2}$, Jennifer E.\ Andrews$^{2,3}$, Peter Milne$^{2}$, \newauthor Charles D. Kilpatrick$^{4}$, and Dan Milisavljevic$^{1,5}$ \\
  $^{1}$Department of Physics and Astronomy, Purdue University, 525 Northwestern Ave, West Lafayette, IN 47907, USA \\
    $^{2}$Steward Observatory, University of Arizona, 933 N. Cherry
  Ave., Tucson, AZ 85721, USA \\
  $^{3}$ Gemini Observatory/NSF’s NOIRLab, 670 N. A’ohoku Place, Hilo, HI 96720, USA\\
  $^{4}$Center for Interdisciplinary Exploration and Research in Astrophysics (CIERA) and Department of Physics and Astronomy,\\ Northwestern University, Evanston, IL 60208, USA\\
  $^{5}$Integrative Data Science Institute, Purdue University, West Lafayette, IN 47907, USA}

\begin{document}
\pagerange{\pageref{firstpage}--\pageref{lastpage}} \pubyear{2022}
\maketitle
\label{firstpage}
\begin{abstract}
We present a series of ground-based photometry and spectroscopy of the superluminous Type~IIn supernova (SN) ASASSN-15ua, which shows evidence for strong interaction with pre-existing dense circumstellar material (CSM). Our observations constrain the speed, mass-loss rate, and extent of the progenitor wind shortly before explosion. A narrow P~Cygni absorption component reveals a progenitor wind speed of $\sim$100 km s$^{-1}$.   As observed in previous SNe~IIn, the intermediate-width H$\alpha$ emission became more asymmetric and blueshifted over time, suggesting either asymmetric CSM, an asymmetric explosion, or increasing selective extinction from dust within the post-shock shell or SN ejecta.  Based on the CSM radius and speed, we find that the progenitor suffered extreme eruptive mass loss with a rate of 0.1-1 M$_\odot$ yr$^{-1}$ during the $\sim$12 years immediately before the death of the star that imparted $\sim$ 10$^{48}$ erg of kinetic energy to the CSM. Integrating its $V$-band light curve over the first 170 days after discovery, we find that ASASSN-15ua radiated at least 3$\times$10$^{50}$~erg in visual light alone, giving a lower limit to the total radiated energy that may have approached 10$^{51}$ erg. ASASSN-15ua exhibits many similarities to two well-studied superluminous SNe~IIn: SN~2006tf and SN~2010jl.  Based on a detailed comparison of these three, we find that ASASSN-15ua falls in between these two events in a wide variety of observed properties and derived physical parameters, illustrating a continuum of behavior across superluminous SNe~IIn. 
\\

\end{abstract}



\section{INTRODUCTION}

Type IIn supernovae (SNe) are identified by relatively narrow H emission lines in their spectra \citep{filippenko1997_opticalspecofSN,schlegel1990}, which are narrower than the broad P Cygni profiles typically seen in normal Type II-P events.  Whereas the broad lines in normal SNe II-P arise from emission by the recombining fast ($\sim$10$^4$ km s$^{-1}$) SN ejecta, the narrow emission in SNe~IIn is typically a superposition of ``narrow" ($\sim$10$^2$ km s$^{-1}$) and ``intermediate-width" ($\sim$10$^3$ km s$^{-1}$) components.   The narrow emission component traces the dense photoionized gas in the pre-shock circumstellar material (CSM), and the intermediate-width emission can arise from either electron scattering (more common in early spectra) or from the accelerated post-shock gas (typically seen at later times). See \citet{smith17} for a review of interacting SNe. 

The progenitors of SNe IIn are believed to undergo extreme mass loss events immediately prior to explosion. Observed mass-loss rates, inferred from optical light curves of SNe IIn and narrow/intermediate-width H$\alpha$, are of order 0.01-1 M$_{\odot}$ yr$^{-1}$ or more \citep{smith14,smith17}, whereas steady radiation-driven stellar wind mass-loss rates are expected to be $\la$10$^{-4}$ M$_{\odot}$ yr$^{-1}$ \citep{smith14,smithowocki2006_contdriveneruptions}. This suggests that an episodic or eruptive mechanism drives the pre-SN mass loss to form dense CSM around SNe~IIn, and comparisons have been made to the giant outbursts of luminous blue variables (LBVs) and SN imposters \citep{smith11_lbv,vandyk05}.  Indeed, there are now several cases where a SN IIn was preceded by an observed eruption in the few years immediately before the SN  \citep{smith10_09ip,fraser13,smith14,ofek14,bilinski15,elias-rosa16}.\footnote{Note that we are not counting cases where a claimed precursor event may have been the SN itself before the onset of CSM interaction, and we are not including the first detection of a pre-SN event in the case of SN 2006jc's precursor \citep{pastorello07} because it was a Type Ibn, not a Type IIn event.}  Sources of this eruptive mass loss may arise from instabilities associated with advanced stages of nuclear burning, including the pulsational pair instability, wave driving, or other nuclear burning instabilities  \citep{qs12,sq14,fuller17,fr18,woosley07,woosley_2017_PPISN,ws22,sa14,renzo20},  or violent LBV-like binary interaction events and pre-SN mergers \citep{sa14,smith2018_etacar_lightechoes,smith22,schroder20}. 

SNe~IIn exhibit wide diversity in terms of luminosity, inferred CSM mass, radial distribution of the CSM (and hence, timing of the mass loss before explosion), and asymmetry \citep{smith17}, but all require pre-SN mass-loss that is much stronger than what normal steady winds of massive stars can provide \citep{smith14}.  Among SNe IIn, the most extreme cases that produce the highest luminosity from CSM interaction are the so-called super-luminous supernovae (SLSNe) that can reach absolute magnitudes $\leq -21$ mag in the $R$-band \citep{galyam_2012_SLSNe}.

\begin{figure*} 
    \centering
    \includegraphics[width=\textwidth]{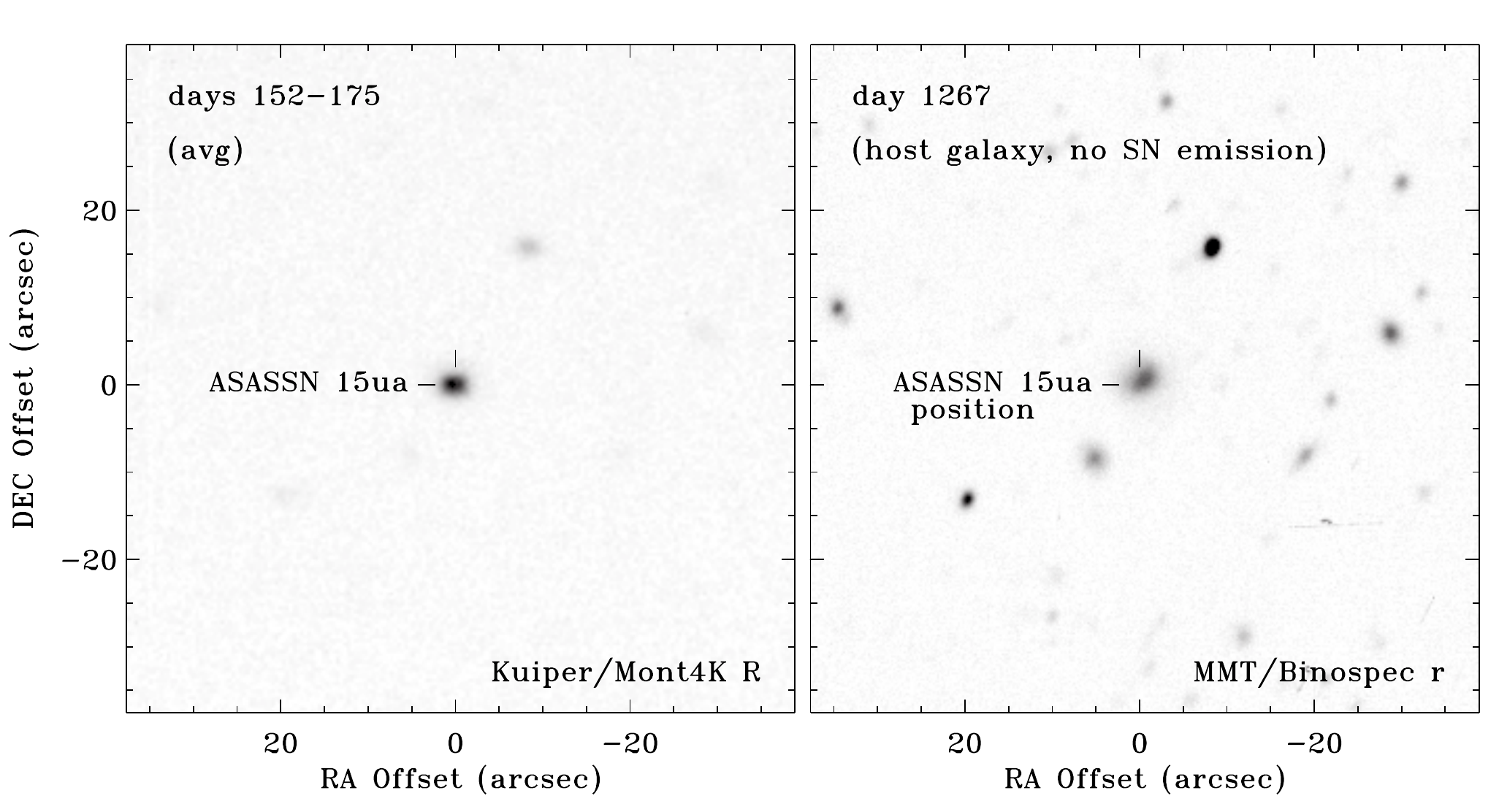}
    \caption{Images of the SN location with and without SN light.  {\it Left:} Kuiper/Mont4k $R$-band image of ASASSN-15ua with the SN location identified.  This is an average of 3 images taken between days 152 and 175 after discovery, when SN light still dominated the observed flux.  {\it Right:}  MMT/Binospec $r$-band image of the same field at late times after the SN faded, showing its location in its host galaxy GALEXASC J133454.49+105906.7 as measured in the Kuiper image in the left panel.}
 \label{fig:image}
 \end{figure*}

The extreme luminosity of SLSNe IIn arises because of the ability of CSM interaction to efficiently convert SN ejecta kinetic energy into radiation, rather than because of extreme explosion energy \citep{sm07,vanmarle10,ci11}.  In other words, the unusual property of SLSNe IIn is related to CSM close to the star, rather than a highly unusual SN explosion.  Most SLSNe are consistent with a normal (i.e. 10$^{51}$ erg) explosion energy, and even the most extreme cases like SN~2006gy may only require an explosion energy of a few $\times 10^{51}$ erg \citep{smith10_06gy}. On the other hand, SLSNe without narrow H lines in their spectra \citep{quimby11} may derive their extreme luminosities from a different mechanism such as energy injection from a magnetar \citep{maeda07,woosley10,kb10}.

Interacting SNe may greatly contribute to the dust budgets of high redshift galaxies \citep{gall14_10jl_dust,bevan_2019_sn2005ip_dust,niculescuduvaz_2022_dust_ccsne}. There are several observational signatures of new dust formation in SNe: (1) infrared (IR) excess, (2) increased fading rate of the optical flux, and (3) a progressive and systematic blueshifting of emission line profiles, which is attributed to post-shock or ejecta dust formation that blocks line emission from the redshifted side of the explosion. An interacting SN’s shock consists of the forward shock, the reverse shock, and a cold dense shell (CDS) that forms in between them \citep{chugai01,smith2008_06tf}. Dust may form in this region due to efficient radiative cooling of post-shock gas collapsing into a thin, dense, and clumpy layer  \citep{smith08_06jc,vanmarle10}. Several H-rich interacting SNe, including SN 2010jl \citep{smith11_10jl, gall14_10jl_dust}, SN 2017hcc \citep{smith2020_17hcc}, and SN 2005ip \citep{smith2009_05ip,bevan_2019_sn2005ip_dust} have shown clear signatures of dust formation, most notably with their similar blueshifted line profiles. Dust masses for SN 2010jl and SN 2005ip have been estimated to be 0.005-0.01M$_\odot$ \citep{bevan_2020_sn2010jl_dust} and 0.1M$_\odot$ \citep{bevan_2019_sn2005ip_dust}, respectively.

In this paper, we discuss the SLSN~IIn event ASASSN-15ua (also known as PS16ub), which was first identified with a $V$-band magnitude of 16.9 mag by  \citet{masi2015_atelfor15ua} on 2015 December 12 (MJD = 57368, UT dates are used throughout this paper). ASASSN-15ua is located 0.95 arcsec south and 1.40 arcsec east from the center of the host galaxy GALEXASC J133454.49+105906.7 \citep{masi2015_atelfor15ua}. Its redshift of $z$ = 0.057 $\pm$0.001 indicates a distance of 256.3 Mpc, assuming H$_0$ = 69.6 km s$^{-1}$ Mpc$^{-1}$, $\Omega_M = 0.286$, $\Omega_\Lambda = 0.714$ \citep{cosmology_calc}. It was discovered on December 12, 2015 (MJD = 57368), while the first detection can be traced back to December 9, 2015 with a \emph{V}-band magnitude of 17.5 mag. We correct for a foreground Milky Way and Host extinction of A$_V=0.178$ mag, using \emph{R$_V$}=3.1 \citep{schlaflyfinkbeiner2011}, Milky Way $E(B-V) = 0.0247$ mag, and host $E(B-V) = 0.0331$ mag (see section \ref{section: na 1 d absorp}, $E(B-V)_{tot} = 0.0578$ mag). The absolute magnitude at discovery then becomes \emph{$M_V$}=$-$20.32 mag. 
ASASSN-15ua’s spectra showed strong narrow emission lines attributed to CSM interaction at early times, indicating that it was a SN IIn event \citep{challis2015_specclassificationsofSN}, and our continued spectral series shows narrow H emission persisting throughout its evolution.

\begin{center}\begin{table*}
\begin{minipage}{\textwidth}
\scriptsize
\centering
\caption{Super-LOTIS and Kuiper Photometry } 
\begin{tabular}{@{}lccccc}\hline\hline

JD	    &   	B (mag)               &	V (mag)	                &	R (mag)	                &	I (mag)   & Instrument 	            	\\	\hline
2457372.21	&	17.51		(0.04)	&	16.95		(0.02)	&	16.73	(0.03)	&   16.41		(0.03)	&   SUPER-Lotis\\	
2457374.22	&	17.48		(0.06)	&	16.89		(0.04)	&	16.72	(0.07)	&	…	            	&   SUPER-Lotis\\	
2457375.22	&	…	    	        &	16.97    	(0.02) &   16.73		(0.02)	&   16.30		(0.04)	&   SUPER-Lotis\\	
2457376.22	&	17.57		(0.07)	&	17.01		(0.04)	&	16.71		(0.06)	&	…	            	&   SUPER-Lotis\\	
2457385.22	&	…           	    &	17.01  	     (0.02) &	16.75	(0.03) &   16.46		(0.04)	&   SUPER-Lotis\\	
2457388.20	&	17.55		(0.11)	&	17.07		(0.04)	&	16.83		(0.04)	&	…	            	&   SUPER-Lotis\\	
2457389.22	&	…   	            &	17.02	     (0.01)	&	16.80	(0.01)	&   16.44		(0.02)	&   SUPER-Lotis\\	
2457390.20	&	17.76		(0.04)	&	17.04		(0.02)	&	16.83		(0.02)	&	…	            	&   SUPER-Lotis\\	
2457391.21	&	…   	           	&	17.06		(0.01)	&	16.82		(0.01)	&   16.45	(0.02)	&   SUPER-Lotis\\	
2457405.22	&	17.81		(0.05)	&	17.15		(0.02)	&	16.89		(0.03)	&	…	            	&   SUPER-Lotis\\	
2457411.21	&	18.06		(0.15)	&	17.10     	(0.05)	&	16.98		(0.06)	&	…	            	&   SUPER-Lotis\\	
2457414.20	&	17.98		(0.15)	&	17.17     	(0.09)	&	16.91		(0.09)	&	…	            	&   SUPER-Lotis\\	
2457423.20	&	18.05		(0.05)	&	17.35		(0.03)	&	17.10		(0.04)	&	…	            	&   SUPER-Lotis\\	
2457426.20	&	18.18		(0.05)	&	17.34		(0.02)	&	17.08		(0.03)	&	…	               	&   SUPER-Lotis\\	
2457429.20	&	18.07		(0.04)	&	17.38		(0.02)	&	17.11	(0.02)	&   16.81  	(0.03)	&   SUPER-Lotis\\	
2457432.20	&	18.18		(0.04)	&	17.35		(0.02)	&	17.11		(0.02)	&   16.76		(0.03)	&   SUPER-Lotis\\	
2457435.20	&	18.22		(0.05)	&	17.38		(0.03)	&	17.13		(0.03)	&   16.78		(0.03)	&   SUPER-Lotis\\	
2457458.09	&	18.30		(0.06)	&	17.57		(0.04)	&	17.22		(0.05)	&   17.93		(0.29)	&   SUPER-Lotis\\
2457520.92	&	18.82 (0.12)	&	17.99 (0.05)	&	17.67 (0.04)	    &   …   & Kuiper\\
2457534.36	&	18.74 (0.07)	&	17.98 (0.07)	&	17.68 (0.08)	    &   …   & Kuiper\\
2457543.12	&	19.06 (0.26)	&	18.08 (0.06)	&	17.83 (0.05)	    &   …   & Kuiper\\
2458857.20	&	20.00		(1.46)	&	18.42		(1.47)	&	…	                &	…   	        	&   SUPER-Lotis\\	
2458873.07	&	…           	&	…	               	&	19.21		(0.09)	&	17.44		(0.44)	&   SUPER-Lotis\\		\hline

\label{tab: superlotis}
\end{tabular}
\end{minipage}\end{table*}

\end{center}

We place ASASSN-15ua in context with the well-studied and seemingly distinct SNe IIn: SN 2006tf and SN 2010jl. The progenitors to both objects has been claimed to be LBV-like stars, but since LBV eruptions are an observed phenomenon without a well understood physical cause, this does not directly address the underlying mechanisms for the pre-SN mass loss. 
Characterizing the diversity of SNe IIn and their progenitors can help constrain the final stages of a massive star's life. We present our observations in Section \ref{section:obs}. The light curve and spectral evolution are discussed in Section \ref{section:results}. Section \ref{section:discussion} reviews our main results, assesses the CSM interaction strength and CSM mass requirements, and discusses them in the context of other luminous CSM interaction-powered SNe. A consistent theme is that in many respects, we find ASASSN-15ua to have observed characteristics and inferred physical properties intermediate between the two well-studied SLSNe~IIn events SN~2006tf  \citep{smith2008_06tf} and SN~2010jl \citep{smith11_10jl,smith2012_10jl,zhang12_10jl,gall14_10jl_dust,jencson2016_10jl}.

\begin{figure*}
    \centering
    \includegraphics[width=.95\textwidth]{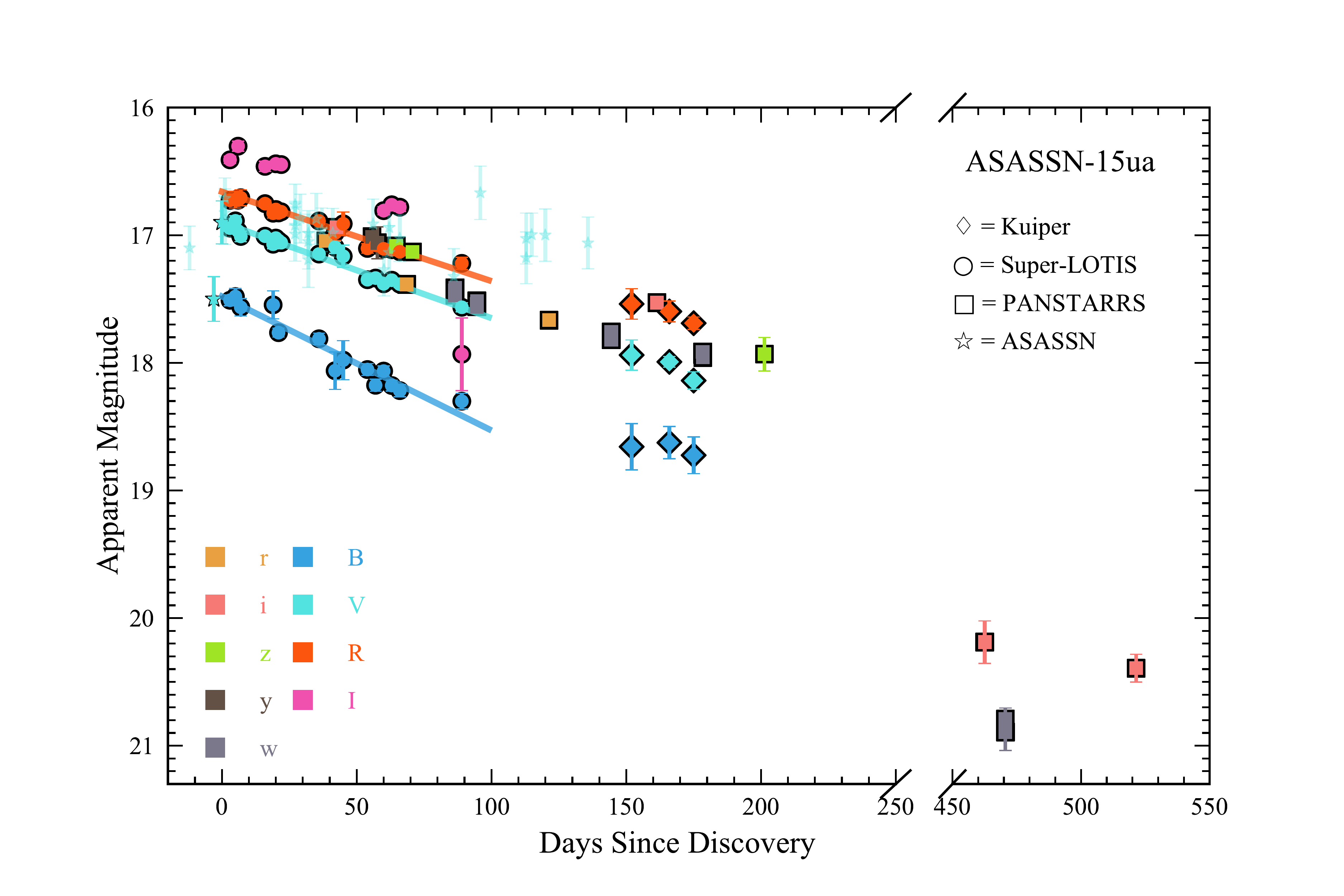}
    \caption{\emph{BVRI} Photometry from Super-LOTIS, ASASSN, and Kuiper (see Table \ref{tab: superlotis}) and \emph{grizwy} photometry from PANSTARRS1. These photometry are not corrected for extinction. Days since discovery is displayed because the time of explosion and time of peak are unknown; this is adopted for the remainder of the paper. The bulk of available photometry is in the first 170 days, which is shown on the left, while $i$ and $w$ band photometry are the only filters  included at late times, shown at right. Diamond markers indicate Kuiper $BVR$ data, circular markers indicate SL $BVRI$ data,  squares indicate PANSTARRS $grizyw$, and light stars indicate ASASSN $V$ mag (light blue). Slopes have been applied to $BVR$ photometry to indicate time evolution in the first 100 days.}
    \label{fig:long app mag}
\end{figure*}

\section{OBSERVATIONS}\label{section:obs}
\subsection{Photometry}

We obtained optical photometry of ASASSN-15ua in \emph{BVRI} filters using the robotic 0.6m Super-LOTIS (Livermore Optical Transient Imaging System; \citealt{williams2008slotis}) telescope located on Kitt Peak.  Standard reductions, including flat-fielding and bias subtraction, were carried out using procedures described by \citet{kilpatrick2016}. Aperture photometry using Landolt catalog in IRAF was performed on the reduced images. Table \ref{tab: superlotis} lists the apparent magnitudes for the Super-LOTIS data. 


We also measured optical \emph{BVRI} photometry of ASASSN-15ua in images obtained with the Mont4k CCD Camera \citep{fontaine14_mont4k} mounted on the 1.5m Kuiper Telescope on Mt. Bigelow, Arizona. Flat fielding and bias subtraction were performed using standard image calibration procedures. Template subtraction was performed on the Kuiper images by the High Order Transform of Psf ANd Template Subtraction code\footnote{\hyperlink{https://github.com/acbecker/hotpants}{https://github.com/acbecker/hotpants}} with pre-explosion Pan-STARRS1 images \citep{tonry2006GPC}. Aperture photometry using Landolt catalog in IRAF was performed on the reduced images.

We also retrieved publicly available \emph{grizyw} photometry from Pan-STARRS1 (PS1, \citealt{PS1}; \citealt{ps1_database}) 3Pi Steradian Survey's Gigapixel Camera \citep{tonry2006GPC}. 
\D{PS1 photometry are PSF magnitudes from template subtracted images using zeropoints in the nightly processed 3Pi images, which are tied to the calibration in \citet{schlafly2010_panstarrs} and \citet{ps1_astrometric_calib}.}

One late-time (Day 1267, June 1, 2019) $r$-band image of ASASSN-15ua was obtained using the imaging mode of Binospec \citep{fabricant2019binospec}, mounted  on the 6.5m MMT Observatory located on Mt. Hopkins, Arizona.  This image is shown in the right panel of Figure \ref{fig:image}.  At this late phase, it appears that the SN has faded beyond detection limits, as there is no residual flux in the template subtracted image. The very faint dwarf galaxy host of ASASSN-15ua is similar to that of SN~2006tf \citep{smith2008_06tf}, and relatively faint dwarf hosts seem to be a common property among SLSNe~IIn \citep{stoll2011_10jl}. 

The multiband photometry of ASASSN-15ua are shown in Figure \ref{fig:long app mag}. We take day zero to be the discovery confirmation date , MJD = 57368, 2015 December 12. The data shown in Fig. \ref{fig:long app mag} have not yet been corrected for extinction.


\begin{figure*}
    \centering
    \includegraphics[width=\textwidth]{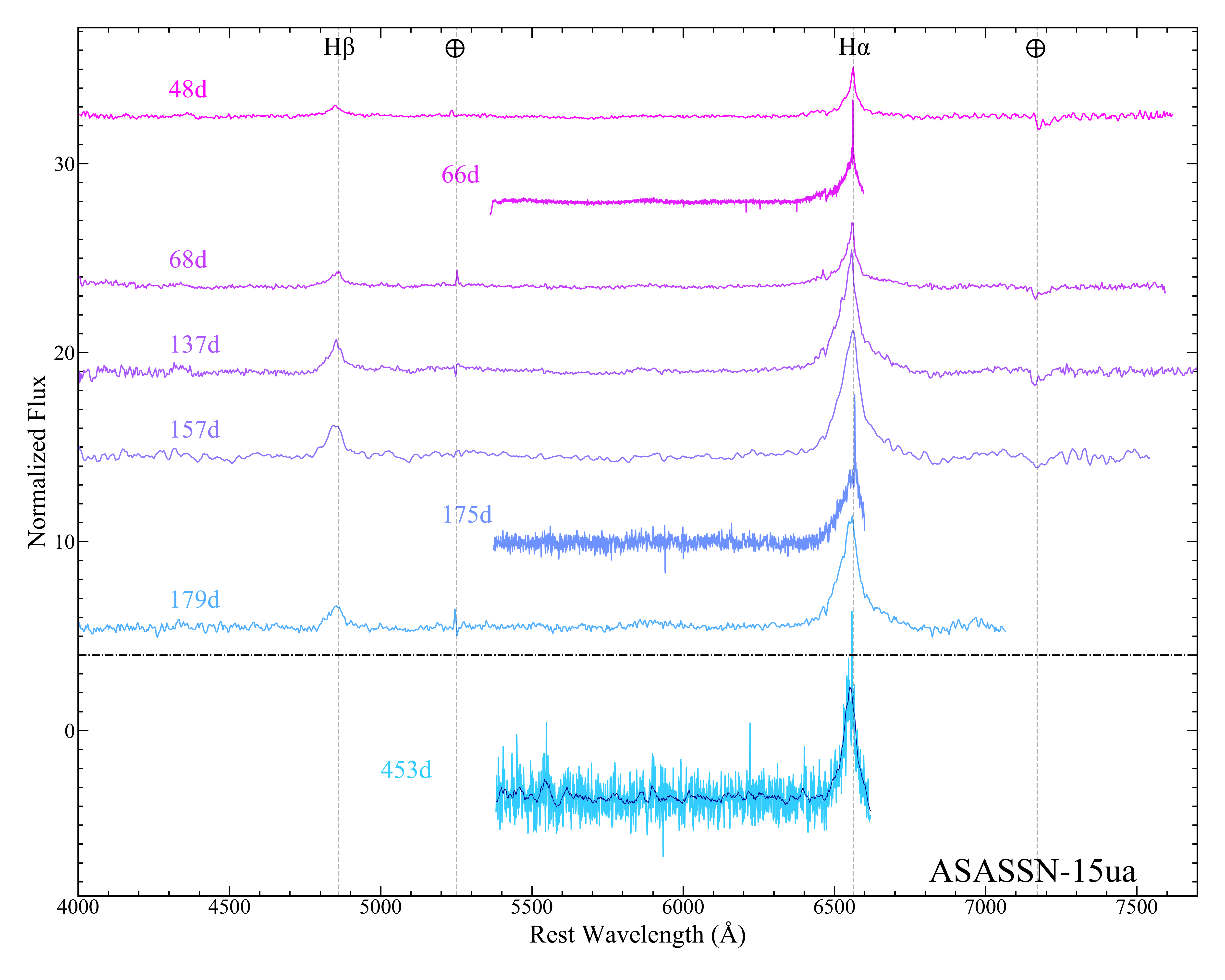}
    \caption{Visual wavelength, continuum-normalized spectra of ASASSN-15ua from Bok and MMT (see Table~\ref{tab:specobs}). The spectra on days 175 and 453 have been smoothed with 3 pixel bins. For day 453, an even further smoothed version of the same spectrum has been overlaid in dark blue.}
    \label{fig:full spec}
\end{figure*}

\begin{figure}
    \centering
    \includegraphics[width=\columnwidth]{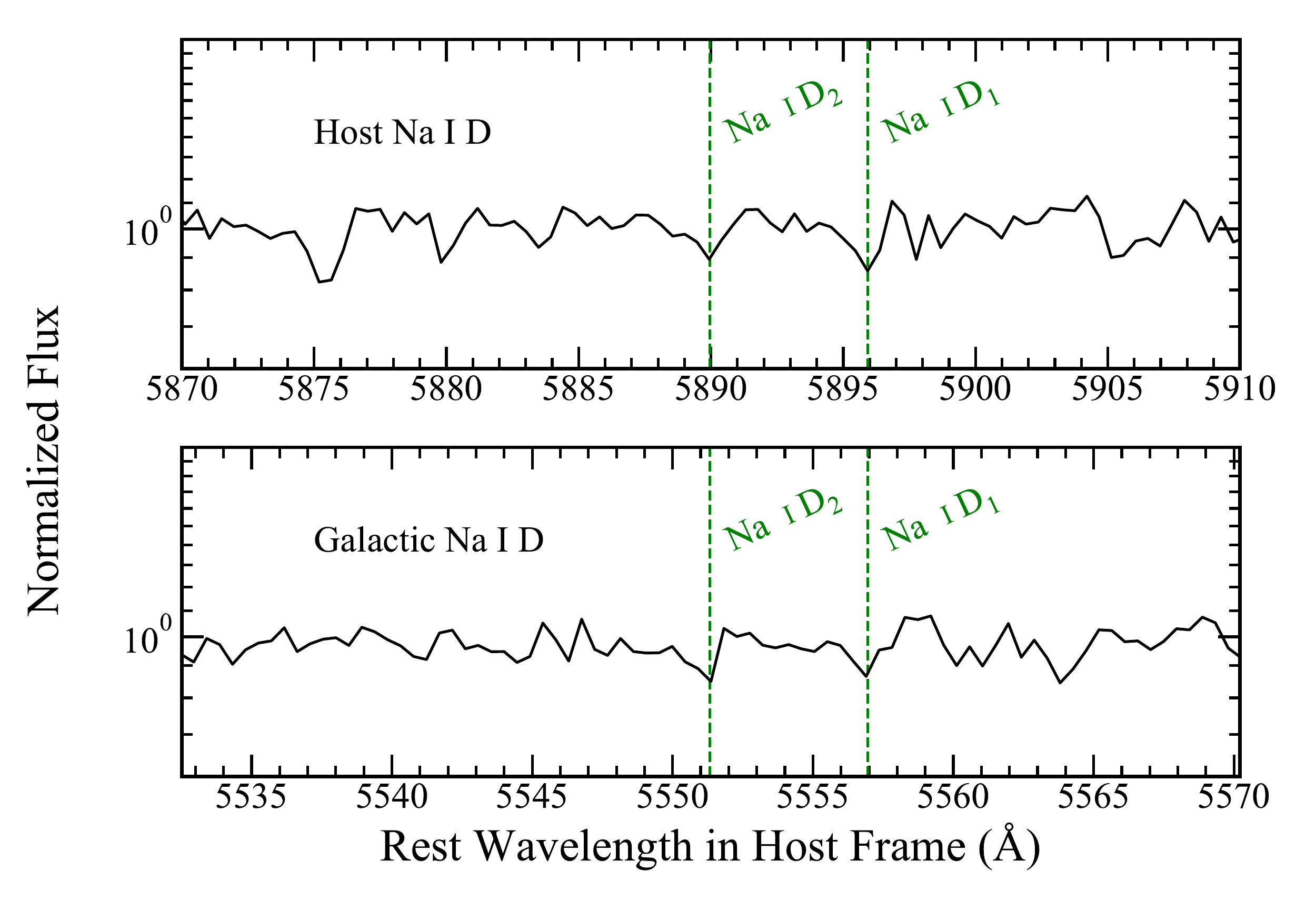}
    \caption{Na {\small I} D doublet in the Day 66 MMT Spectrum. In both panels, wavelengths have been corrected to show the rest wavelength at the redshift of ASASSN~15ua, and as a result, the Milky Way absorption components appear significantly blueshifted.}
    \label{fig:na 1 D}
\end{figure}

\begin{figure} 
    \centering
    \includegraphics[width=\columnwidth]{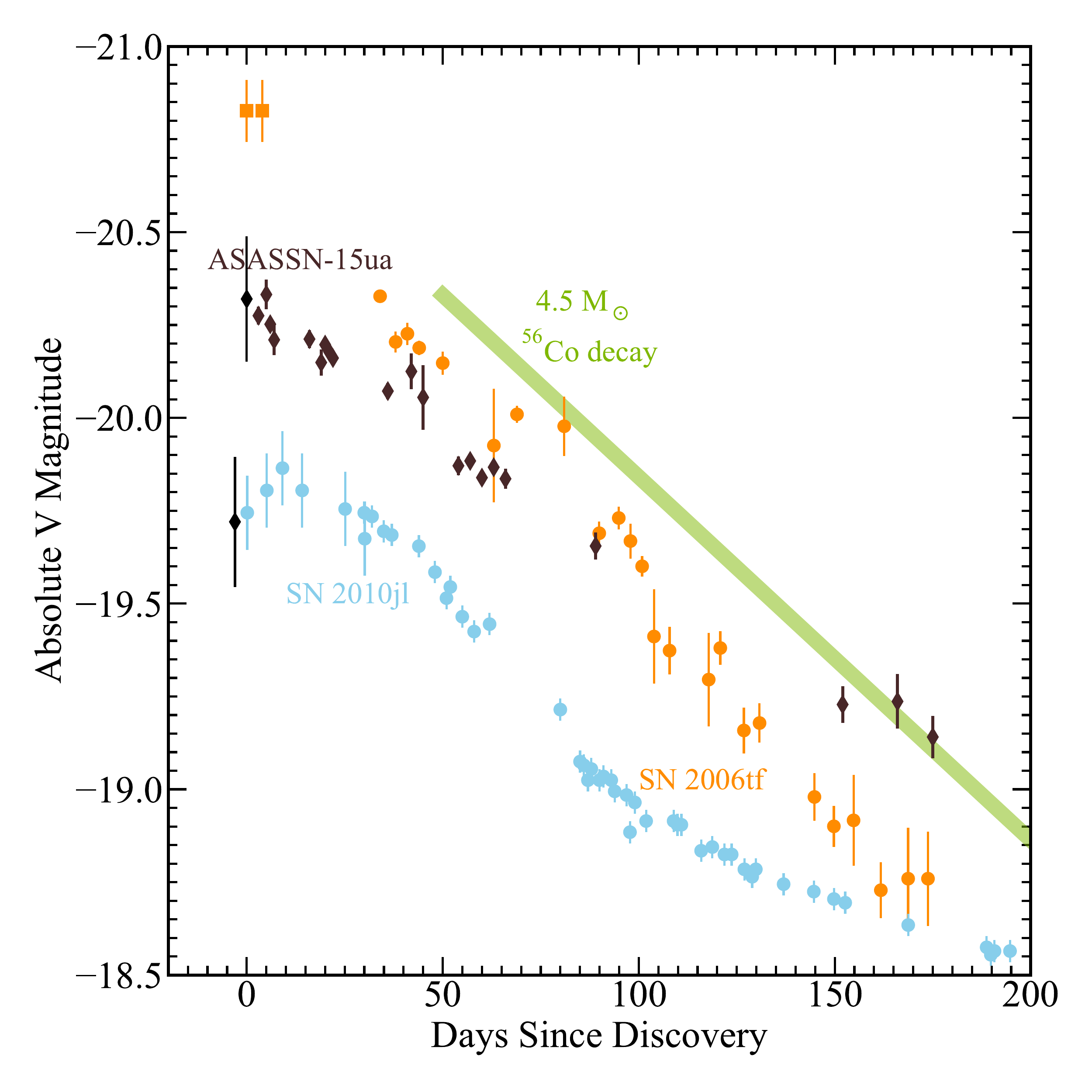}
    \caption{Absolute $V-$band magnitudes for IIn SLSNe: ASASSN-15ua, SN2006tf \citep{smith2008_06tf}, and SN2010jl \citep{Fransson_2014_SN2010jl,stoll2011_10jl}.  SN~2006tf and SN~2010jl have been adjusted to absolute magnitude adopting $E(B-V)$ values and distances described in the text. The orange squares are the unfiltered discovery observations of SN~2006tf. A green slope corresponding to the radioactive decay rate consistent with 4.5M$_\odot$ of $^{56}$Co is shown for comparison.} 
    \label{fig:lightcurve}
\end{figure}

\subsection{Spectra}
During the 170 days after discovery, we obtained eight optical spectra of ASASSN-15ua, with observations summarized in Table \ref{tab:specobs} and the spectra plotted in Figure~\ref{fig:full spec}. Five of the spectra were from the Bok 2.3m Telescope using the Boller \& Chivens Spectrograph (B\&C Spectrograph; \citealt{angel1979bcspec}); the epochs for these spectra are 48, 68, 137, 157, and 179 days after discovery, with a 300 lines mm$^{-1}$ grating, slit width of 1.5", wavelength coverage of 3700 to 8000 {\AA} (3200 to 7500 {\AA} in rest wavelengths), and approximate resolution of 8 {\AA} (365 km s$^{-1}$ at H$\alpha$).   The other three spectra had higher spectral resolution, and were obtained with the 6.5-m MMT using the Blue Channel Spectrograph, covering the epochs of 66, 175, and 453 days, obtained with the 1200 lines mm$^{-1}$ grating that yielded a wavelength coverage of 5750 to 7000 {\AA} (5400 to 6600 {\AA} in rest wavelengths), and approximate resolution of 1.45 {\AA} (66 km s$^{-1}$ at H$\alpha$). The slit was oriented along the parallactic angle for all MMT and Bok observations.

\begin{table*}\begin{center}\begin{minipage}{\textwidth}
\scriptsize
\centering
\caption{Spectroscopic Observations}
\begin{tabular}{@{}lcccccl}\hline\hline
Date  & MJD  &Day$^a$ &   Tel./Inst.  &   Range  & Exposure &   Resolution \\
    & &   &   & {\AA} & s & {\AA} (km s$^{-1}$) \\ \hline
2016 Jan 29  &  57416.43   &   48  & Bok/B\&C  &   3900-8100   &   600  &   $\approx$8 (365)   \\
2016 Feb 16  &  57434.47   &   66  & MMT/BC    &   5750-7000   &   600  &  1.45 (66) \\
2016 Feb 18  &  57436.46   &   68  & Bok/B\&C  &   3700-8000   &   1200 &   $\approx$8 (365)  \\ 
2016 Apr 27  &  57505.40   &   137 & Bok/B\&C  &   4000-8300   &   900  &   $\approx$8 (365)  \\
2016 May 17  &  57525.36   &   157 & Bok/B\&C  &   3700-8000   &   600  &   $\approx$8 (365)   \\ 
2016 Jun 4   &  57543.21   &   175 & MMT/BC    &   5740-7000   &   600  &  1.45 (66) \\
2016 Jun 8   &  57547.32   &   179 & Bok/B\&C  &   3800-7600   &   600  &   $\approx$8 (365)  \\ 
2017 Mar 9   &  57821.46   &   453 & MMT/BC    &   5740-7000   &   1200 &  1.45 (66)  \\
\hline
$^{a}$Day 0 is taken as the date of discovery.
\label{tab:specobs}
\end{tabular}
\end{minipage}\end{center}
\end{table*}

\subsection{Host Na {\small I} D Absorption}
\label{section: na 1 d absorp}

We marginally detect Na {\small I} D absorption due to the host galaxy in the higher resolution spectrum taken 66 days after maximum light (see Figure \ref{fig:na 1 D}). We measure the equivalent widths of Na {\small I} D$_1$ and D$_2$ to be 0.122$\pm 0.031$ {\AA} and 0.162$\pm 0.010$ {\AA}, respectively. Using Equation 8 in \citet{poznanski2012_Na1D}, we calculate a host $E(B-V) = 0.033 \pm 0.018$ mag. Adding this to the foreground Milky Way extinction quoted earlier, we adopt a total $E(B-V) = 0.058$ mag.


\section{RESULTS}\label{section:results}

\subsection{Light Curve}
\label{section: results light curve}
Photometry obtained with Super-LOTIS, PS1, ASASSN, and Kuiper are shown in Figure \ref{fig:long app mag}. Slopes are are obtained by fitting to the $BVR$ photometry for the first 100 days. The corresponding light curve decay rates are 0.011, 0.0075, and 0.0069 mag d$^{-1}$, respectively. While the decay rate of the $B-$band light curve is similar to the radioactive decay of $^{56}$Co, the light curves in other bands decay more slowly, indicating that the bolometric luminosity decay is slower than $^{56}$Co.  This suggests that radioactive heating is not the only power source for the observed luminosity. The observed $B-V$ color gets redder at a rate of 0.0031 mag d$^{-1}$.

Figure \ref{fig:lightcurve} shows the absolute $V$ magnitude light curve of ASASSN-15ua covering its first 170 days after discovery. The two well-studied SNe IIn, SN~2006tf and SN~2010jl, are included for comparison. Note that the light curves of these two objects have been corrected for both distance and Milky Way extinction, following values in \citet{smith2008_06tf} and \citet{smith11_10jl}. For SN~2006tf we adopted $E(B-V)= 0.027$ mag \citep{smith2008_06tf}and $d=308$ Mpc \citep{quimby_2007b_06tf}, and for SN~2010jl we adopted $E(B-V) = 0.024$ mag \citep{jencson2016_10jl}  and $d=46.5$ Mpc (host $z = 0.0107$, NED\footnote{The NASA/IPAC Extragalactic Database (NED) is operated by the Jet propulsion Laboratory, California Institute of Technology, under contract with the National Aeronautics and Space Administration.}). ASASSN-15ua is corrected for host and Milky Way extinction as described in section \ref{section: na 1 d absorp}.   At early times, ASASSN-15ua's light curve resembles that of SN~2006tf in terms of its peak luminosity and rate of decline, although ASASSN-15ua appears to fade more slowly. In contrast, the plateaued decline of SN~2010jl slows after day 100 and maintains a high luminosity until about day 300. This implies that the radial distribution of CSM around ASASSN-15ua more closely resembles the compact configuration of SN~2006tf. 

Integrating the $V$-band light curve over the 170 days after discovery, we measure a total radiated energy of $E_{\rm rad} \sim 3\times$10$^{50}$ erg, which is not unusual for SLSNe. We did this by integrating a Gaussian Process Regression model \citep{garretson_gpr_2021} of the V-band photometry from 3 to 174d. Note that this is an underestimate of the true value of $E_{\rm rad}$ because ASASSN-15ua was discovered at peak luminosity or after. The early rise to peak was largely missed, the late-time tail is not included, and we have made no bolometric correction (BC) to the optical light curve (there may be a significant BC at early times when the effective temperature is hotter).  With these considerations, the total $E_{\rm rad}$ is likely to be around 10$^{51}$ erg.

Figure \ref{fig:lightcurve} includes a thick green slope indicating the expected decline rate for radioactive decay of 4.5 M$_\odot$ of $^{56}$Co. Day zero for SN~2006tf was determined using unfiltered discovery observations that measured M$_r=-20.7$ mag. Day zero for SN~2010jl is the day of discovery.
\reftwo{The V-band light curve of ASASSN-15ua declines more slowly than the trend expected for radioactive decay, making it unlikely that its light curve is dominated by a large mass of $^{56}$Co.  Instead, CSM interaction powers the luminosity, given the spectral signatures of ASASSN-15ua, as is the case with most SLSNe IIn.}
We use the luminosity and radiated energy of ASASSN-15ua to constrain its total CSM mass and progenitor mass-loss rate below, after we consider expansion speeds of the SN and CSM indicated by spectra.

\subsection{Low-resolution Spectra}
In our analysis of spectral evolution, we will primarily consider the spectra obtained from the Bok telescope, as the MMT spectra have limited wavelength range. ASASSN-15ua has such a high redshift that the H$\alpha$ line resides near the edge of the  observed wavelength range used for the MMT observations, unfortunately truncating the red wing of the line.  The higher resolution MMT spectra are therefore used primarily to investigate the narrow P-Cygni lines arising from pre-shock CSM (see Section 3.3).

\subsubsection{Line Profile Decomposition}
\label{section: line profile decomp}
\begin{figure*}
    \centering
    \includegraphics[width=\textwidth]{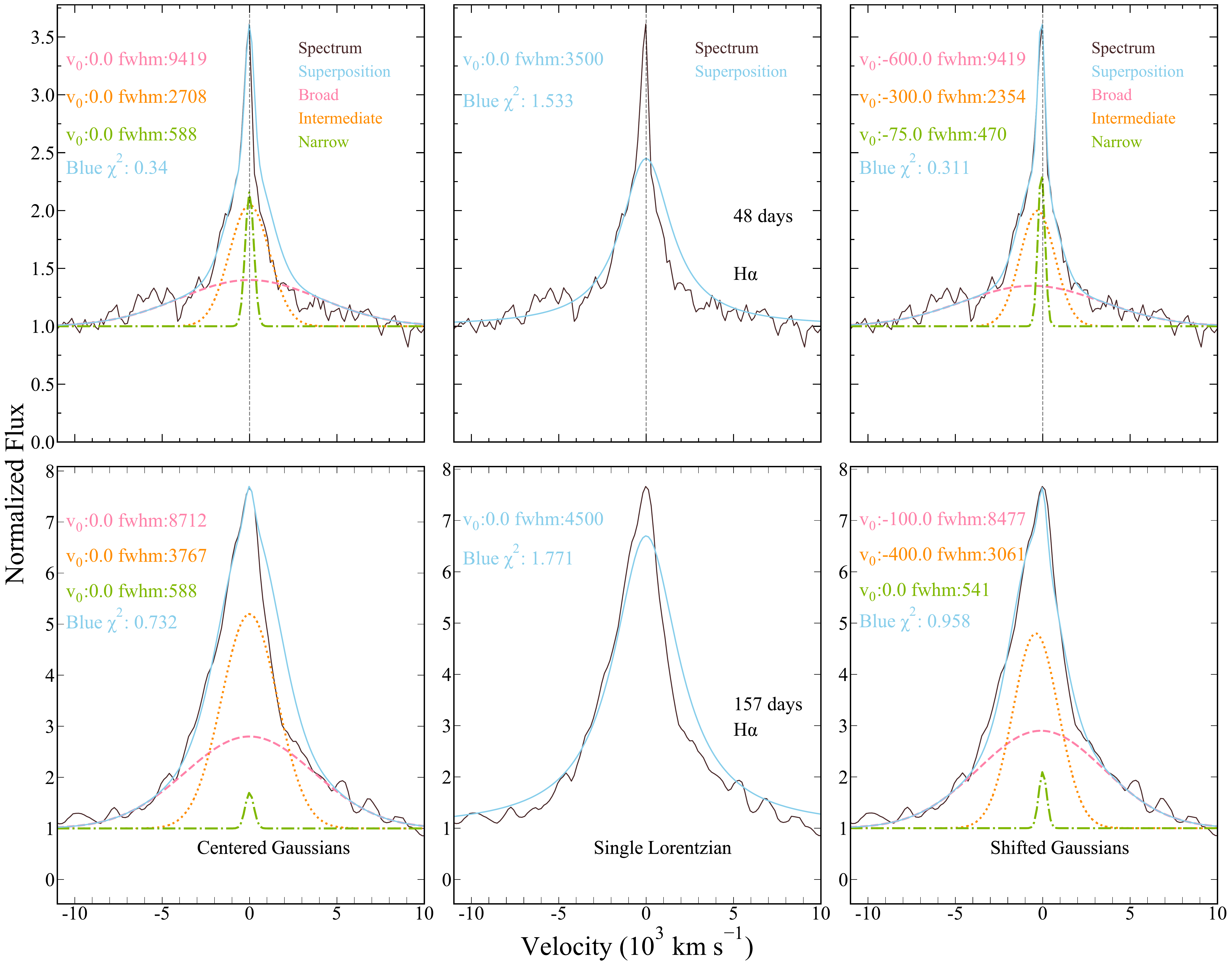}
    \caption{Various functions plotted against the H$\alpha$ profile in the Bok spectra observed on days 48 and 157. $Left:$ 3 Gaussians of varying width centered at 0 km s$^{-1}$, the center of the narrow H$\alpha$ component. Gaussians have been fit to the blue side of the line, since the profile is asymmetric. $Center: $ A single Lorentzian centered at 0 km s$^{-1}$. $Right: $ 3 Gaussians of varying width shifted with respect to the center of the line. The three Gaussians represent the broad, intermediate, and narrow components of the complex line profiles. The $\chi^2$ of each function is listed; this value represents only the goodness of fit between the data and function on the blue side of the line.} 
    \label{fig:profile fits}
\end{figure*}

The profile of Balmer lines in SNe IIn is complicated. At early times, electron scattering in the dense CSM broadens the narrow component into an intermediate-width profile with a Lorentzian shape \citep{chugai01}. This profile usually evolves into a sum of two (or three) Gaussians, as the optical depth of the CSM drops and emission from the post shock gas (and sometimes also SN ejecta) is directly observable \citep{smith2008_06tf,smith17}.

We compared three possible functions to the observed H$\alpha$ profile of ASASSN-15ua for each epoch of our Bok spectra \D{(shown in Figure \ref{fig:profile fits} are only days 48 and 157).} The first function is comprised of three Gaussians centered at 0 km s$^{-1}$ with widths corresponding to the narrow, intermediate, and broad-width components. Due to progressive blueshifting of the line, these Gaussians were only compared to the shape of the blue side. The second is a single Lorentzian representing an electron-scattered intermediate-width component. The last is three Gaussians where the centroids were allowed to shift to arbitrary nonzero values. This function aims to describe the whole shape of the complex line profile. 

Figure \ref{fig:profile fits} shows these comparisons for days 48 and 157 (similar comparisons were made for all the Bok spectra but are not shown here). It is clear from the center column that the shape of H$\alpha$ is never Lorentzian after 48 days. Across all epochs, the Lorentzian cannot describe the red side of the line. By day 48, most of the CSM is evidently no longer optically thick to electron scattering. We cannot well constrain the time of explosion, so the SN may be older than 48 days at that time, and hence, it is likely that none of our spectra captured the early-time behavior.

The H$\alpha$ line is better described by three Gaussians. The broad component of both the centered and shifted Gaussians are generally similar in width for each epoch. The intermediate-width component of the centered Gaussians are systematically broader than their shifted counterparts, since they ignore the red wing of the line. The narrow components of both Gaussians do not show a clear trend, but it may be difficult to capture the narrow, 100 km s$^{-1}$, component because it is unresolved (resolution $\sim$ 600 km s$^{-1}$ at H$\alpha$) and blended with broader components. No obvious trend is seen in the zero-point velocity of any Gaussian curves in the shifted column, except that all non-zero shifts are to the blue. We suggest that it makes more physical sense for the centered Gaussian curves to describe the shape of the line, acknowledging that the red side of the line may be suppressed due to extinction from dust in the ejecta or CDS. 

As these fits are not unique, nearly identical superpositions of the Gaussians could be produced with components of slightly different widths and strengths. What is shown here attempts to employ components as narrow and as strong as possible, guided by widths of components often observed in SNe IIn.

The intermediate-width and broad components help guide our estimates for the expansion speed of the post-shock CDS and the SN ejecta, respectively. We adopt a shock speed of 2,500 km s$^{-1}$. The speed of the SN ejecta is observed to be of order 4,000-5,000 km s$^{-1}$. We will adopt CSM speeds from the day 66 P Cygni profile, discussed later (see Section \ref{section: p cyg results}).


\subsubsection{Line Asymmetry and Evolution}

\label{section: line asymm and evo}

\begin{figure}
    \centering
    \includegraphics[width=\columnwidth]{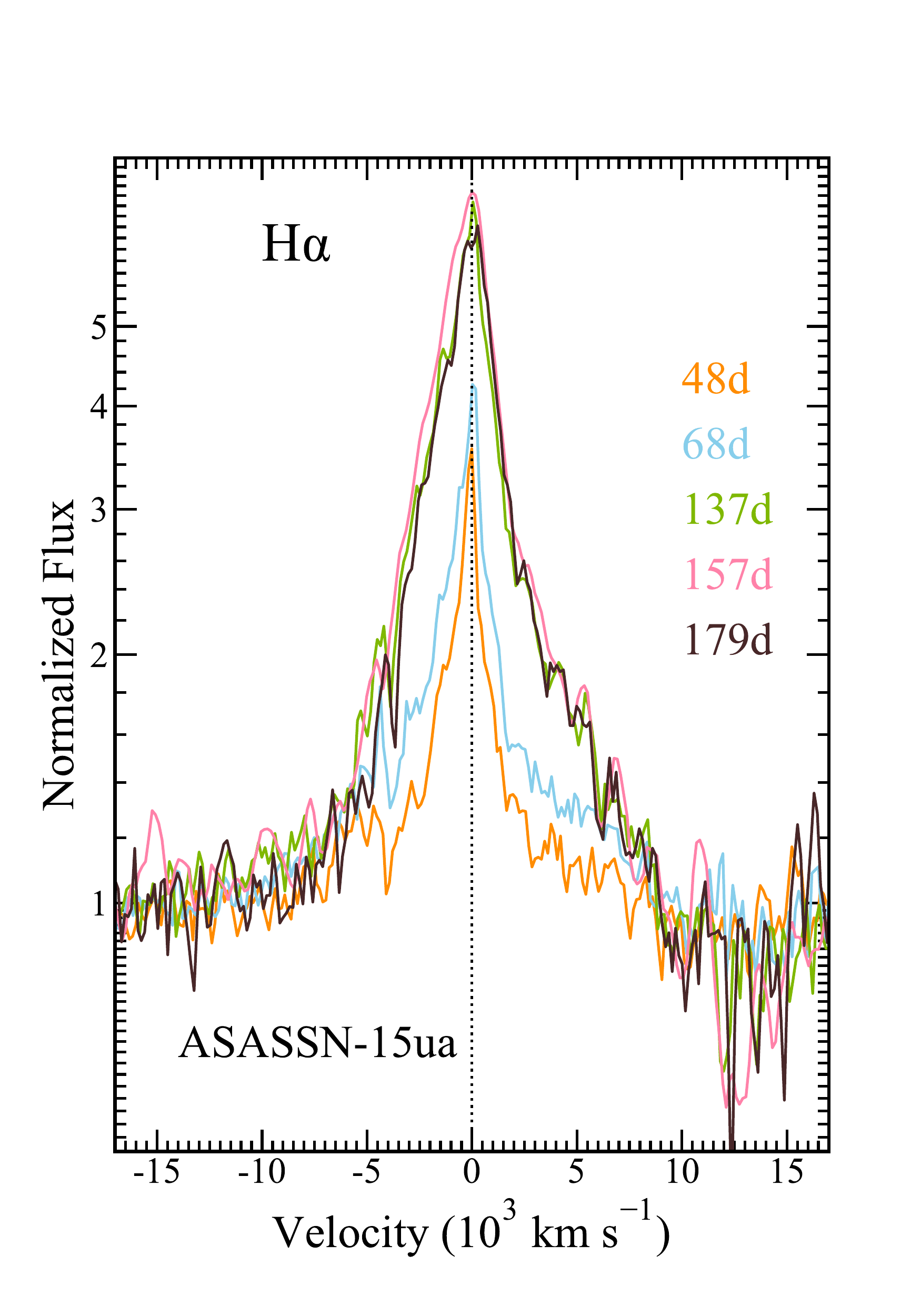}
    \caption{Evolution of the H$\alpha$ line-profile shape seen in continuum-normalized Bok spectra of ASASSN-15ua. Note that the line core (a mix of  the intermediate-width and unresolved narrow components at this resolution) becomes broader with time, but the blue wing stays approximately the same.}
    \label{fig:H alpha evolution}
\end{figure}

\begin{figure}
    \centering
    \includegraphics[width=\columnwidth]{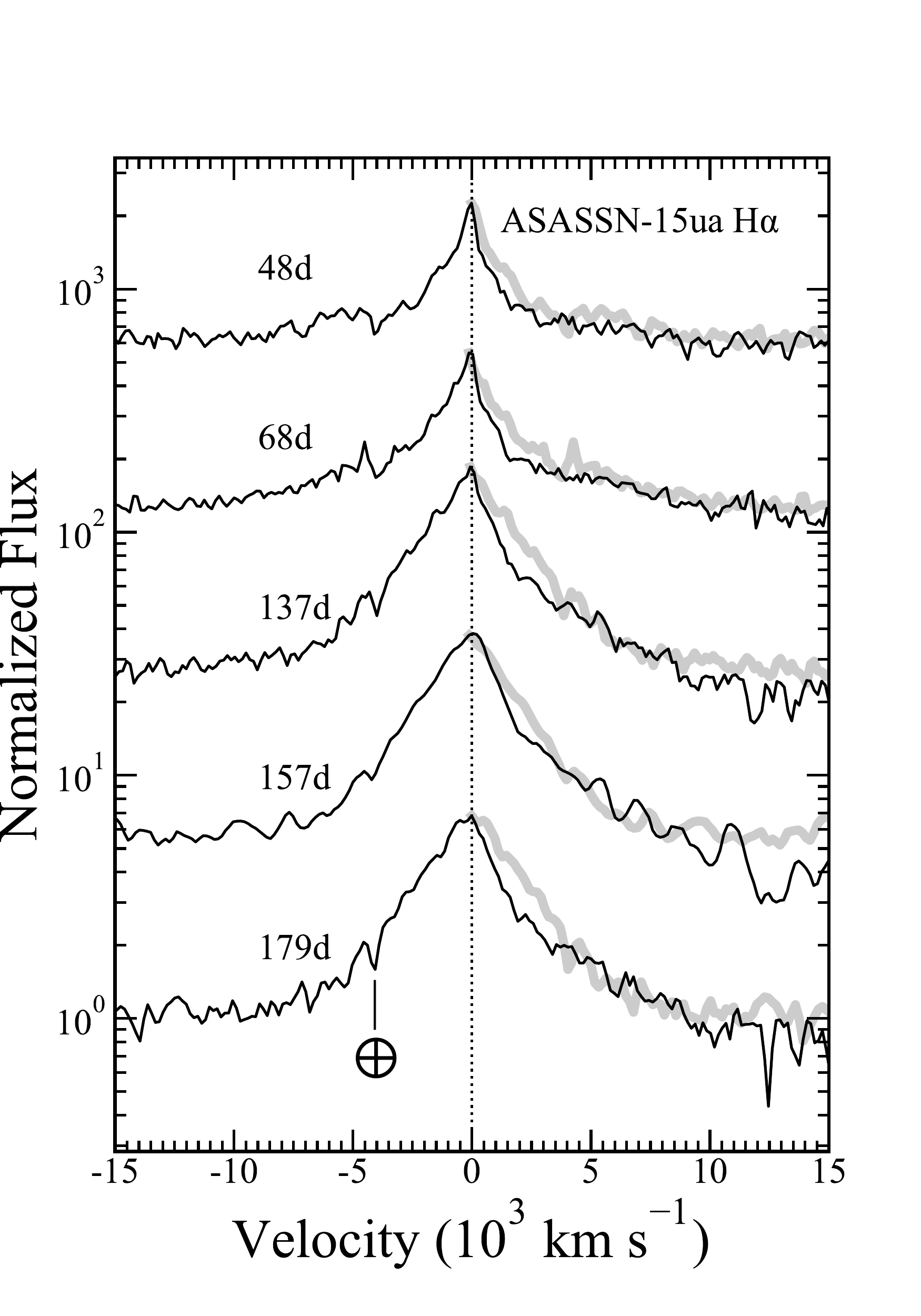}
    \caption{The same H$\alpha$ line profiles as in Figure~\ref{fig:H alpha evolution} but separated vertically. \D{This plot shows the blue-shifted wing of the line reflected over (shown in a thicker light grey line) to redshifted velocities.} There is a consistent flux deficit on the observed red wings, even though the narrow line peak is at zero velocity.}
    \label{fig:uaspecreflec}
\end{figure}


Figure \ref{fig:H alpha evolution} emphasizes the intermediate-width H$\alpha$ profile on days 48, 68, 137, 157, and 179. The spectra are continuum-normalized.  The classic Lorentzian electron-scattering shape of the emission, often seen in early-time spectra of SNe IIn, is not seen in ASASSN-15ua (see section \ref{section: line profile decomp}). Over time, we see the three characteristic widths of H$\alpha$ emission in SNe IIn. All three components strengthen with time compared to the continuum. The widths of the narrow and broad component stay roughly constant, while the intermediate component becomes progressively blueshifted (recall that the narrow component is unresolved in these Bok spectra, but see section \ref{section: p cyg results} for a look at narrow component evolution). 

The component spanning $\pm$2000 km s$^{-1}$ accounts for the majority of the flux and dominates the overall appearance of the line profile. There is a sudden widening in the intermediate component of the spectra after 100 days. At each epoch, the blueshifted wing is stronger than the opposite side, as shown in more detail in Figure \ref{fig:uaspecreflec}. This figure shows the full observed H$\alpha$ line profile (black) but also shows a mirrored version of the blue wing of the line, reflected over to the red side to demonstrate what the line profile would look like if it were symmetric.  The blue wing reflected over to redshifted velocities is shown as a lighter, thicker line. Figure \ref{fig:uaspecreflec} clearly demonstrates the asymmetry in the H$\alpha$ profile, but it indicates the flux deficit on the redshifted wing is rather mild in ASASSN-15ua at these early phases.

At early times (48d and 68d), the broad and intermediate-width components are blueshifted, but the broad component becomes symmetric after 100d. The asymmetry only persists in the intermediate-width component. This is clearer in the centered Gaussian curves in Fig \ref{fig:profile fits}. The narrow component is fairly symmetric in all epochs, and the intermediate-width profile is systematically blueshifted, with an average shift of -350 $\pm$ 100 km s$^{-1}$. This could indicate early dust formation primarily in the post-shock gas.

\subsubsection{Line strengths and widths}

\begin{figure*}
    \centering
    \includegraphics[width=\columnwidth]{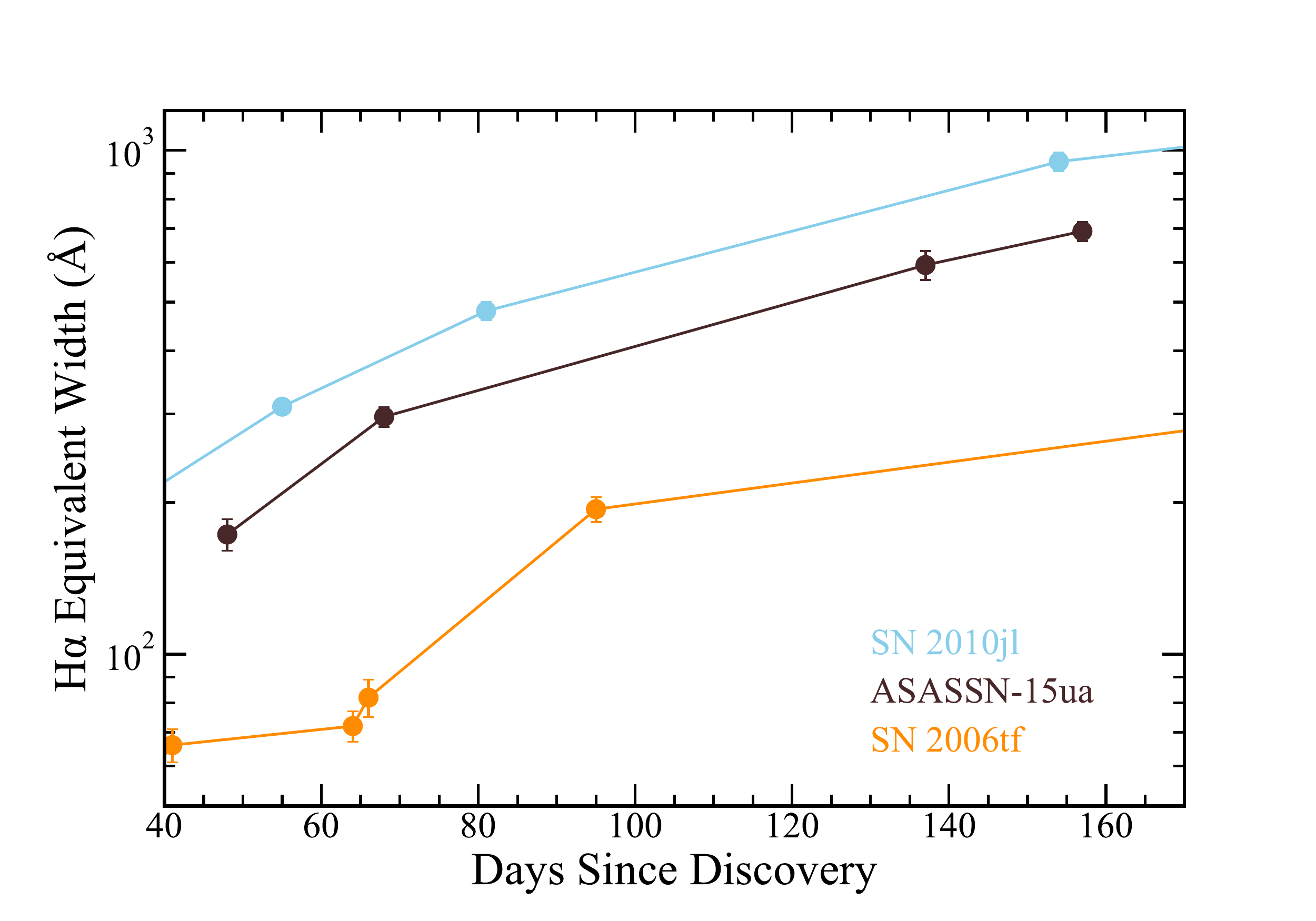}
    \includegraphics[width=\columnwidth]{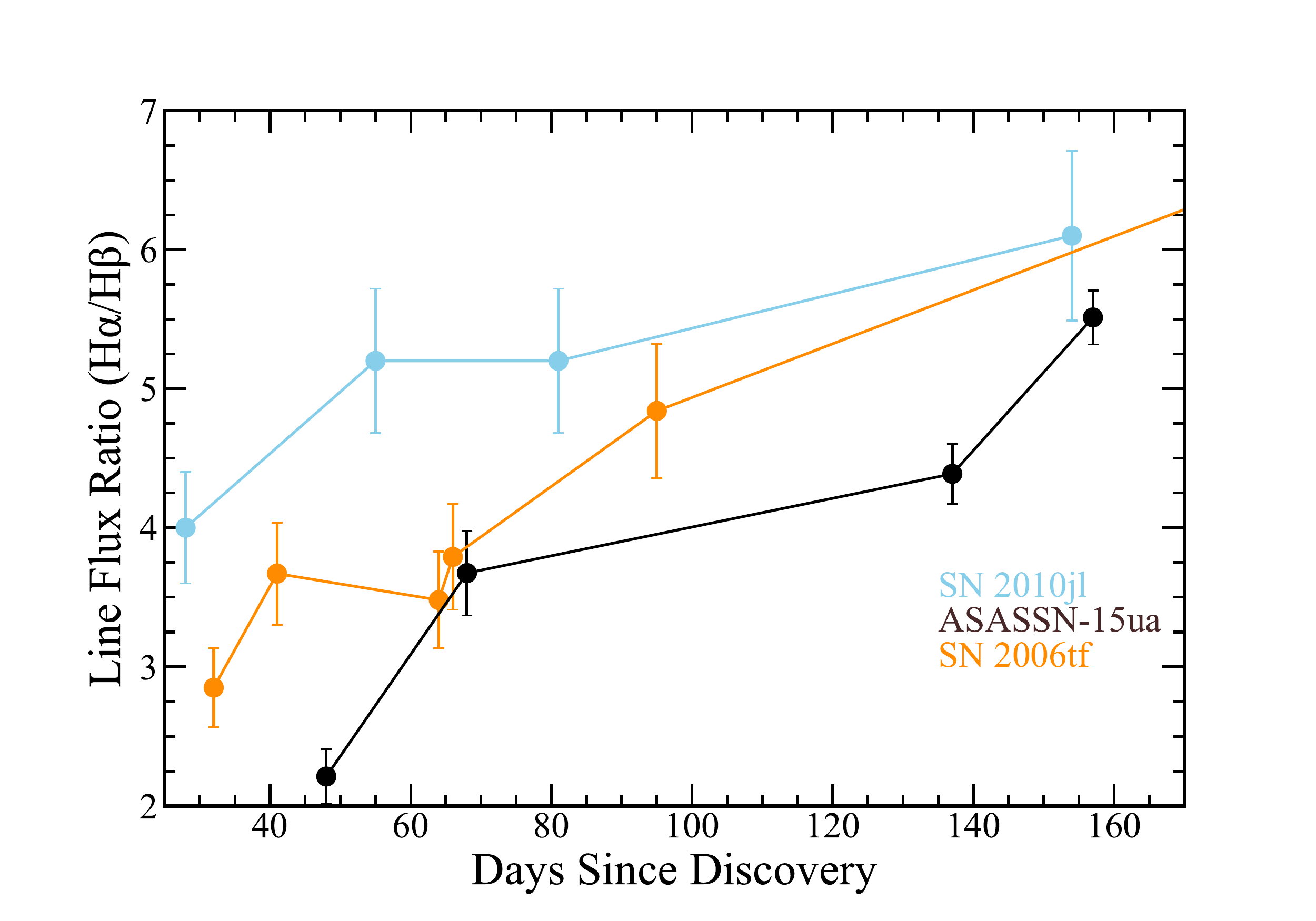}
    \caption{\emph{Left:} Equivalent width (EW) of ASSASN-15ua's H$\alpha$ line, as well as the H$\alpha$ EW of SN 2010jl \citep{jencson2016_10jl} and SN 2006tf \citep{smith2008_06tf}. \emph{Right:} Flux ratio of H$\alpha$/H$\beta$ for ASASSN-15ua, SN2010jl \citep{jencson2016_10jl}, and SN2006tf \citep{smith2008_06tf}. The fluxes and ratios were measured from dereddened spectra, with $E(B-V)$ values for each SN noted earlier. Measurements for day 453 were omitted due to insufficient signal to noise.}
    \label{fig:ew and ratio vs time}
\end{figure*}

\begin{table*}\begin{center}\begin{minipage}{6.2in}
\scriptsize
\centering
\caption{Measurements of Balmer emission lines }

\begin{tabular}{@{}lcccccl}\hline\hline

Date & MJD & Day$^a$ & EW(H$\alpha$)& F(H$\alpha$)&   H$\alpha$/H$\beta$ \\
    &&&{\AA} &   10$^{-14}$erg/s/cm$^{2}$  &  \\ \hline
2016	Jan	29	&	57416.43	&	48	&	-172.84	(12.46)	&	5.02	(0.29)	&	2.21 (0.20)	\\					
2016	Feb	18	&	57436.46	&	68	&	-295.98	(13.11)	&	2.77	(0.17)	&	3.67 (0.30)	\\					
2016	Apr	27	&	57505.40	&	137	&	-592.16	(39.00)	&	4.87	(0.08)	&	4.38 (0.22)	\\			
2016	May	17	&	57525.36	&	157	&	-691.15	(30.46)	&	11.3	(0.3)	&	5.15 (0.19) \\						
2016	Jun	8	&	57547.32	&	179	&	-546.44	(58.56)	&	5.97	(0.11)	&	4.29 (0.12) \\				
\hline
$^{a}$Day 0 is the date of discovery, MJD = 57368.\\
\label{tab:spectra}
\end{tabular}
\end{minipage}\end{center}
\end{table*}

Our measurements of the time evolution of the equivalent width (EW) of H$\alpha$ are plotted in Figure \ref{fig:ew and ratio vs time}, where the H$\alpha$ EW of ASASSN-15ua is again compared to values for SN~2006tf and SN~2010jl. ASASSN-15ua is clearly intermediate between the two other SNe IIn, as the line strength increases with time. These values are positive for emission and include the total line emission from the narrow, intermediate, and broad-width line components. The H$\alpha$/H$\beta$ line flux ratio is also compared to the ratio observed for SN2006tf and SN2010jl. 

Since EW is a measure of line flux relative to the continuum flux, epochs with brighter continuum emission will tend to result in a smaller EW. As the continuum fades through time, the EW of H$\alpha$ increases, as seen in Figure \ref{fig:ew and ratio vs time}.  This is a common property of SNe IIn where the luminosity is dominated by CSM interaction \citep{smith14sn09ip,smith17}. ASASSN-15ua's H$\alpha$ EW evolution is similar in value and slope to SN~2010jl; if the post-shock gas in both objects cools at the same rate, this suggests that the CSM density profiles in ASASSN-15ua and SN~2010jl are similar. SN~2006tf has a lower EW due to additional continuum luminosity from delayed photon diffusion in its opaque shocked shell \citep{smith2008_06tf}.

The line flux ratio between H$\alpha$ and H$\beta$, seen in Figure \ref{fig:ew and ratio vs time} (right panel), shows all three objects increasing at similar rates during the main peak of the light curve.  Note that due to the uncertain explosion epochs of both ASASSN-15ua and SN~2006tf, the horizontal offsets between the three SNe in this plot are uncertain, but ASASSN-15ua and SN~2006tf will only shift to the right, allowing more time to elapse between explosion and discovery. The time of maximum light in SN~2010jl is well-constrained, since its rise was captured.

At early times, the H$\alpha$ and H$\beta$ emission appear to be dominated by recombination after photoionization, with a line ratio close to a value of $\sim$3. At later times, the ratio rises to much larger values, incompatible with recombination emission. ASASSN-15ua evolves similarly to the two comparison SNe IIn but it seems to have a systematically smaller ratio at each epoch. \D{Since all three SNe produce a H$\alpha$/H$\beta$ ratio much larger than 3, the emission is dominated by collisional excitation (see section 5.5 in \citealt{smith2008_06tf} for more information), and the CSM is optically thin.} 


\subsection{Narrow P Cygni Absorption}
\label{section: p cyg results}
Figure \ref{fig:p cygni} shows the relatively high resolution MMT/Bluechannel spectra of ASASSN-15ua on days 66 and 175 after discovery. Both epochs exhibit narrow P Cygni absorption, although the velocity of this absorption changes.  On day 66, the narrow  P Cyg absorption trough minimum is at $-$85$\pm$10 km s$^{-1}$ with a blue edge at $-$100$\pm$10 km s$^{-1}$.  On day 175, however, the narrow P Cyg absorption appears shallower and broader, with a trough centered at roughly $-$95$\pm$10 km s$^{-1}$  and a blue edge at $-$150$\pm$50 km s$^{-1}$.  Judging by the narrow absorption at early times and the narrow emission at both epochs, we adopt a pre-shock CSM expansion speed of 100$\pm$10 km s$^{-1}$ in our analysis. We note, however, the the shift in P Cyg absorption to somewhat higher velocity at the later epoch might indicate a CSM with either a varying outflow speed, or an impulsive mass ejection that created a Hubble-like flow (i.e., where faster ejecta from the same pre-SN eruption have travelled to farther radii in the CSM, and therefore get overtaken at a later epoch by the shock).


\section{DISCUSSION}\label{section:discussion}




\subsection{Mass Loss History}
In this section, we use observed properties of ASASSN-15ua to infer the mass-loss rate of the progenitor and approximate duration of enhanced mass loss prior to explosion. 

\begin{figure}
    \centering
    \includegraphics[width=\columnwidth]{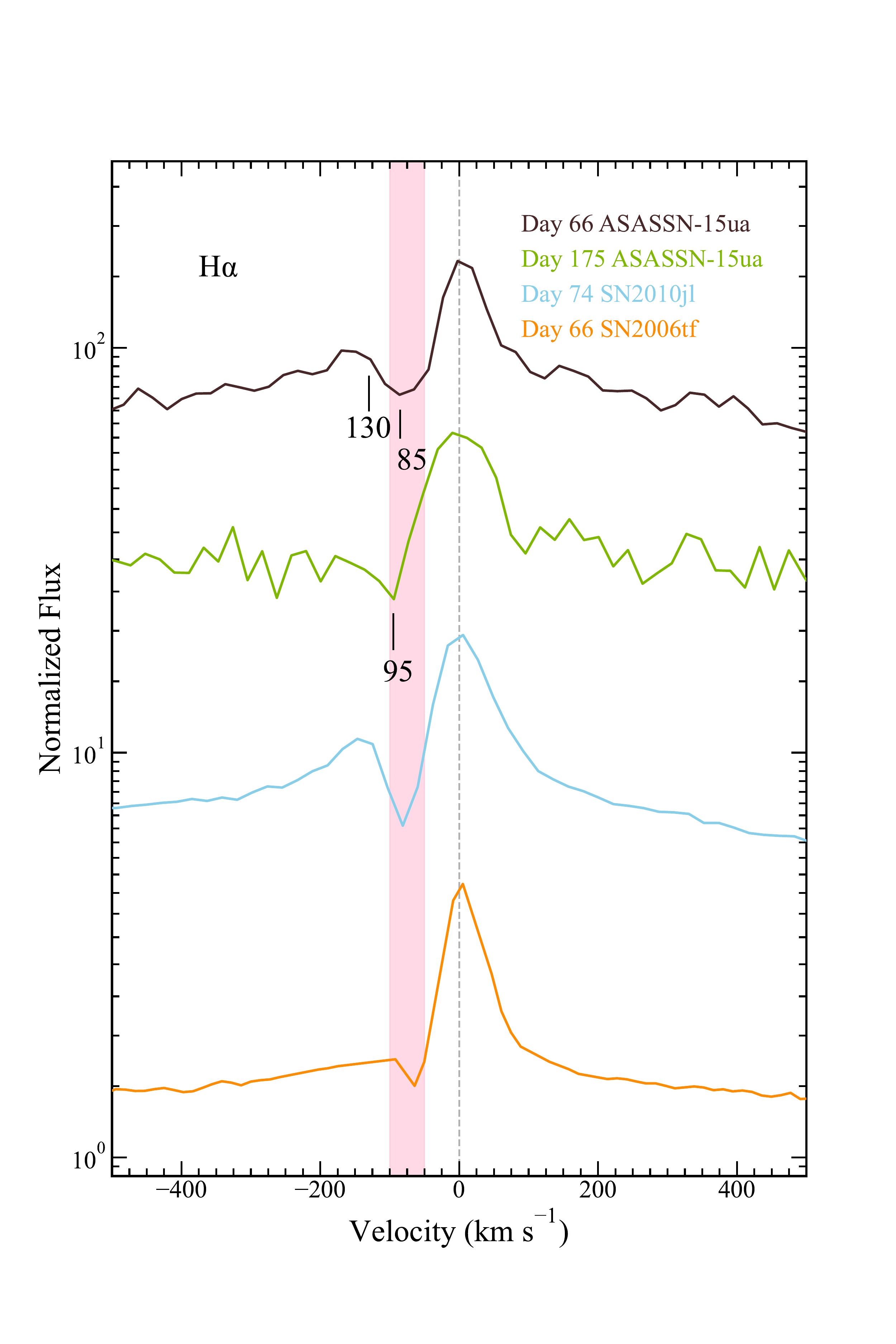}
    \caption{The narrow P Cygni profile of  H$\alpha$ on days 66 and 175 compared to SN 2010jl and SN 2006tf on days 62 and 64, respectively.}
    \label{fig:p cygni}
\end{figure}

We use the narrow emission and absorption as an indicator of the expansion speed for the pre-SN mass loss.  Narrow P Cygni absorption of H$\alpha$ can be seen in the high resolution MMT spectra taken on days 66 and 175. These narrow features are shown in Figure \ref{fig:p cygni} alongside spectra of SN 2010jl and SN 2006tf. The shape of the P Cygni profile is well resolved on day 66, so we take the speed of the absorption's blue edge, $\sim$100 km s$^{-1}$, as the speed of the CSM on that day. A hot pixel or cosmic ray residual was present in the Day 175 MMT spectrum within the apparent P Cygni absorption trough. This was carefully corrected from the 2-D spectrum before extraction, but we nevertheless regard the resulting P Cygni absorption on day 175 with caution. We thus adopt 100 km s$^{-1}$ for the CSM speed at the epoch, assuming constant CSM speed from early to late times. The width of the narrow component on Day 175 appears wider than Day 66, as can be seen from Figure \ref{fig:p cygni}. The CSM speed immediately ahead of the shock may be increasing over time as the forward shock marches outward through faster material at larger radii, hence the apparent acceleration from 85 km s$^{-1}$ to 95 km s$^{-1}$. Nevertheless, we adopt a constant CSM speed of 100$\pm$10 km s$^{-1}$, as this change \D{(10 km$^{-1}$)} in CSM speed would only modify $\dot{M}$ by 10\%, which falls under our uncertainty in the measurement and is far under the spectral resolution, 66 km s$^{-1}$ at H$\alpha$.

For the speed of the forward shock, which is often traced by the intermediate-width component of H$\alpha$ emission from the CDS, we adopt an approximate value of 2,500 km s$^{-1}$. This choice is somewhat slower than the FWHM values of the intermediate-width components at later times in Fig 7, but we chose a somewhat lower value because these lines may be partly blended with the broad component from SN ejecta (note that the intermediate-width component's FWHM value in the left column of Fig 7 increases at late times, even though we might expect the shock to slow down), and lower values are more commonly seen in the shock speeds of other SNe IIn, which are typically between 2,000 and 3,000 km s$^{-1}$ \citep{smith2008_06tf,jencson2016_10jl,smith2020_17hcc}.

Assuming that the observed SN luminosity is dominated by CSM interaction,
the mass-loss rate of the progenitor that would have been required to produce the dense CSM is given by

\begin{align}
\dot{M}_{CSM}&= 2 L \frac{V_{CSM}}{(V_{shock})^3} 
    \label{equation for mass loss rate}
\end{align}

\noindent where Equation \ref{equation for mass loss rate} \citep{smith2008_06tf}\footnote{Note that this quantity is very sensitive to a choice in $v_{shock}$ as $\dot{M} \propto v_{shock}^{-3}$. However, regardless of shock speed within a reasonable range, the calculated mass loss rate is always intermediate to SN~2006tf and SN~2010jl.} calculates the progenitor mass-loss rate as a function of the observed CSM-interaction luminosity ($L$), shock speed ($V_{shock}$), and pre-shock CSM speed ($V_{CSM}$). The luminosity used is the $R$-band magnitude corrected for extinction at each epoch. This estimate assumes that the observed $R$-band luminosity traces 100\% of the CSM interaction luminosity.  Since some of the CSM interaction luminosity may escape at other wavelengths (e.g., X-rays), this value of $L$ --- and hence the corresponding value of $\dot{M}$ --- are lower limits. 

In addition to quantifying the progenitor mass-loss rate, we can determine the approximate time prior to explosion when the CSM was ejected. We can leverage the ratio between $V_{shock}$ and $V_{CSM}$ and $t_{SN}$, the time of observation, assuming constant $V_{shock}$. The consequent equation is as follows:

\begin{align}
    t&=t_{SN}\frac{V_{shock}}{V_{CSM}} \label{time eq}
\end{align}

\noindent where the pre-SN ejection time of the CSM, $t$, is really a lower limit, because the observed value of $t_{SN}$ is taken as the observed time, whereas the true explosion time was earlier than our day zero and because we assume that the shock does not decelerate, which may occur as it encounters CSM. We find that ASASSN-15ua underwent extreme mass loss at a rate of roughly 0.4 M$_\odot$ yr$^{-1}$ for at least 12 years prior to explosion, with an even higher rate of 0.7 M$_\odot$ yr$^{-1}$ within 4.5 years prior to explosion. The SN progenitor ejected a total of at least 6 M$_\odot$ of stellar material into the circumstellar environment, and moving at $\sim$100 km s$^{-1}$, carried a kinematic energy of 6 $\times 10^{47}$ erg. In this model, photons are emitted instantly from the post-shock gas.

\begin{figure}
    \centering
    \includegraphics[width=\columnwidth]{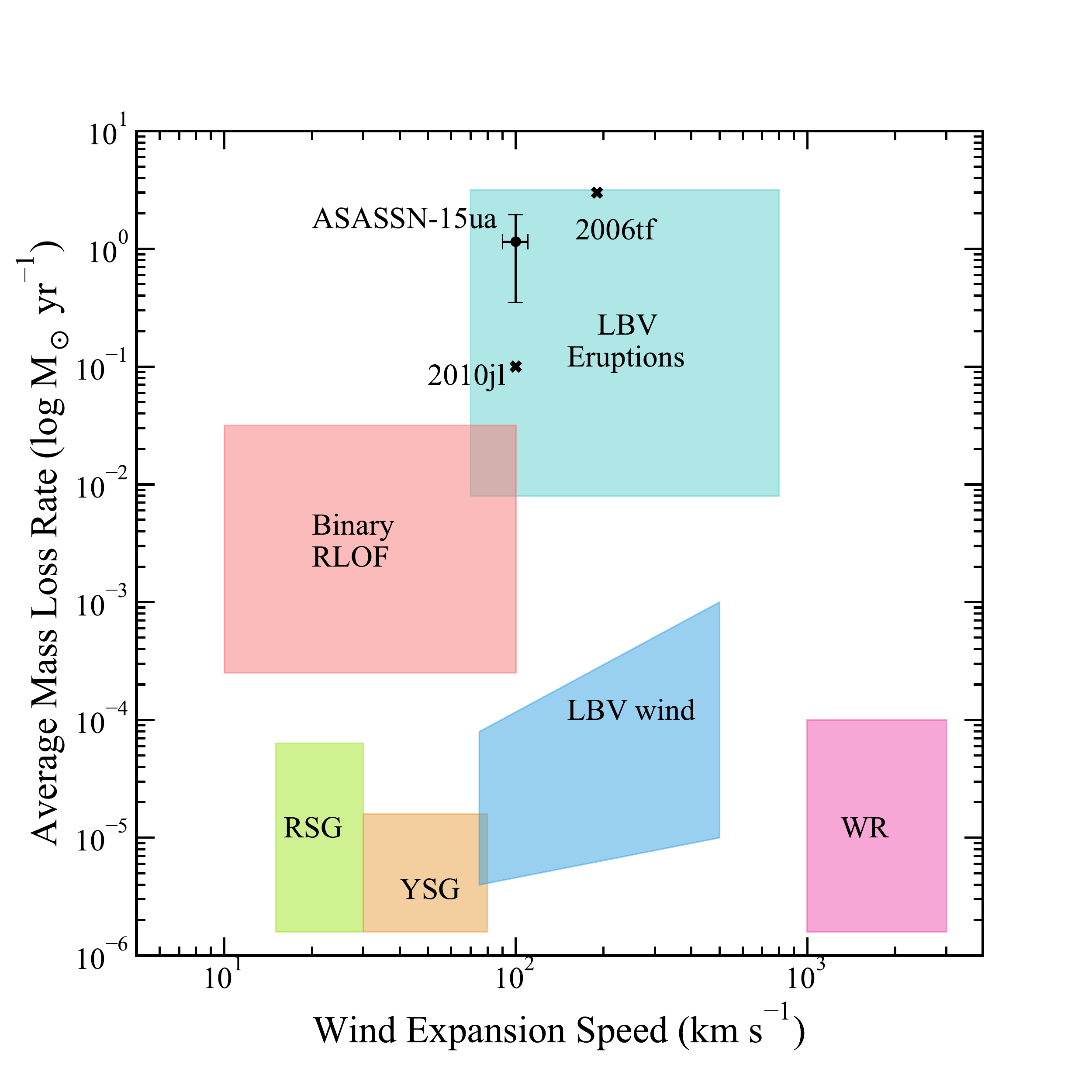}
    \caption{Constraining ASASSN-15ua's progenitor in Mass Loss Rate - Wind Velocity Space. This is adapted from \citet{smith17}. SN2010jl and SN 2006tf have been marked on this plot as well.}
    \label{fig:mass loss wind velocity space}
\end{figure}

The progenitor of SN~2006tf had been experiencing mass loss at a rate of 0.1 M$_\odot$ yr$^{-1}$ in roughly a century before explosion, but then increased  to 2$-$4 M$_\odot$ yr$^{-1}$ in the decade prior to eruption \citep{smith2008_06tf}. The eruptive mass loss of SN~2010jl occurred over about 30 years prior to eruption at a constant 0.1 M$_\odot$ yr$^{-1}$. Figure \ref{fig:mass loss wind velocity space} is adapted from \citet{smith17} and includes the three objects ASASSN-15ua, SN~2006tf, and SN~2010jl. It shows the distribution of mass loss and wind velocity derived from different types of massive stars. One can see that all three objects lay within ranges of LBV giant eruptions but are not consistent with normal stellar winds.

With such high mass-loss rates, the most likely progenitor for ASASSN-15ua is probably an LBV or another object that experienced an LBV-like eruption \citep{smithowocki2006_contdriveneruptions,smith07_06gy,sm07,galyam07_05gl,smith10_09ip}. This progenitor likely had a series of eruptive outbursts or violent binary interaction \citep{sa14}, expelling a large amount of mass into the circumstellar environment. 

An eruptive mechanism would be further validated by a well-resolved P Cygni profile in the day 175 spectrum. If the higher velocity seen in the P Cyg absorption on day 175 is real, that would suggest there is faster pre-shock material along our line of sight.  Since the shock has expanded to larger radii in the CSM at this later time, this may suggest that the pre-SN mass loss was not characterized by a constant velocity wind, but rather an impulsive LBV-like explosion. A similar effect was seen in the case of SN 2006gy, where the increasing CSM speed encountered by the shock at later times outlined a Hubble-like flow, which in turn suggested that the pre-SN mass loss occurred in a single major eruptive burst 8 yr before the SN, rather than a steady wind \citep{smith10_06gy}. Alternatively, the increasing and broader P Cygni absorption may be similar to the case of SN~2017hcc, where higher resolution echelle spectra document the absorption profiles in greater detail, and provide evidence that the changes in line of sight absorption might be related to a bipolar geometry in the CSM shell \citep{smith2020_17hcc}.

\subsection{Line Asymmetry and Possible Dust Formation}
\label{section: discussion dust}
ASASSN-15ua's H$\alpha$ emission becomes progressively more blueshifted with time, as discussed above and illustrated in Figure \ref{fig:profile fits}. The three-Gaussian function centered at 0 km s$^{-1}$ tends to overestimate flux on the red side of the line for each epoch. The asymmetry is most pronounced in the intermediate widths of the lines. Another view of this effect is shown in Figure \ref{fig:uaspecreflec}, where the difference between the reflected blue side (grey) and the red flux (black) grows with time. Note that because of the substantial redshift of ASASSN 15ua, there is a telluric feature near -4,000 km s$^{-1}$, so the difference between the two sides is actually larger than inferred from this figure. 

The asymmetry seen in the line profile is consistent with dust formation, where the newly formed dust in the SN ejecta or in the post-shock CDS causes selective extinction that blocks light from the far side of the SN more than the near side, thus preferentially suppressing redshifted emission in line profiles. Our spectral observations did not capture IR wavelengths, so it is unknown if ASASSN-15ua also had an IR excess. Additionally, the relatively low signal-to-noise ratio for the H$\beta$ line and lack of IR spectra made it difficult to determine if the red-wing suppression exhibits a wavelength dependence, as seen in previous examples like SN~2010jl \citep{smith2012_10jl, gall14_10jl_dust} and SN~2017hcc \citep{smith2020_17hcc}. However, as seen in Figure \ref{fig:compare ha velocity}, which places ASASSN-15ua in context with SN~2006tf and SN~2010jl, all SNe show similar asymmetric spectral evolution. Notably, ASASSN-15ua has only a mild red flux deficit like SN~2006tf, whereas the symmetry seen in H$\alpha$ emission of SN~2010jl is much more extreme. Following the interpretation discussed for several previous SNe, it is likely that dust formation occurred primarily in the post-shock gas, since the majority of the asymmetry occurred in the intermediate width component of the H emission. In ASASSN-15ua, the broad H$\alpha$ component from the fast SN ejecta did not exhibit such a strong deficit of emission on the red wing. 

\D{Dust formation obviously impacts the color of ASASSN-15ua, however, the decay rate of the light curve is heavily dominated by the density profile of the CSM, which is difficult to constrain in this study. Additionally, the quality and sampling of the light curves collected during time periods when spectra show evidence of dust formation is not sufficient for analysis, so we therefore do not use the light curve to infer properties of the dust formed in the CDS.}

In principle, a blueshifted line shape may also be indicative of asymmetric CSM or an asymmetric explosion. The asymmetry would need to be a non-axisymmetric distribution (i.e. one-sided) in order to produce an asymmetric line as observed, with the bulk of the CSM on the near side of the SN.  If the asymmetric lines are due to asymmetric CSM, then one would expect a random orientation of that one-sided CSM due to randomized viewing angles, and we should see roughly equal numbers of blue- and red-shifted asymmetry in SNe~IIn.   While a non-axisymmetric feature is rarely observed as redshifted emission lines \citep{bilinski18}, the blueshifting as observed here is much more common. Moreover, in a SLSN IIn where the luminosity is dominated by CSM interaction, the emission from the fast SN ejecta is likely powered by back-illumination from the forward shock \citep{smith2008_06tf}; this is especially true at later epochs.  However, if the CSM were one-sided, then we would expect to also see blueshifted asymmetry in the fast SN ejecta of ASASSN~15ua, but this is not observed.  And finally, the narrow emission components appear to be centered at zero velocity, arguing against a strongly one-sided distribution of CSM.  We confirm that the CSM around ASASSN-15ua is at least axisymmetric, since the narrow and broad components are symmetric at all times. Although we cannot conclusively rule out aspherical CSM, the progressive blueshifting over time is most likely indicative of dust formation. As in the case of SN~2017hcc \citep{kumar_and_eswaraiah_2019_SN2017hcc_specpol}, spectropolarimetry would have allowed us to further constrain the geometry of the CSM surrounding ASASSN-15ua.

Moreover, ASASSN-15ua shows the classic asymmetric evolution of SNe IIn, as seen in SN~2006tf \citep{smith2008_06tf} and SN~2010jl \citep{smith11_10jl, gall14_10jl_dust}, as well as SN~2017hcc (IIn, \citealt{smith2020_17hcc}) and SN~2006jc (Ibn, \citealt{smith08_06jc}). Dust evidently began to form in the post-shock gas by around $\geq$48 days after explosion; there is some uncertainty is this estimate, since the time of explosion relative to the day of discovery is unknown. SN~2010jl (100d), SN~2017hcc (50-100d), and SN~2006jc (50-75d) formed dust in this region at a similar epoch.

\subsection{An Intermediate Case Between SN2006tf and SN2010jl}
\label{section: int btwn 06tf and 10jl}
\begin{figure}
    \centering
    \includegraphics[width=\columnwidth]{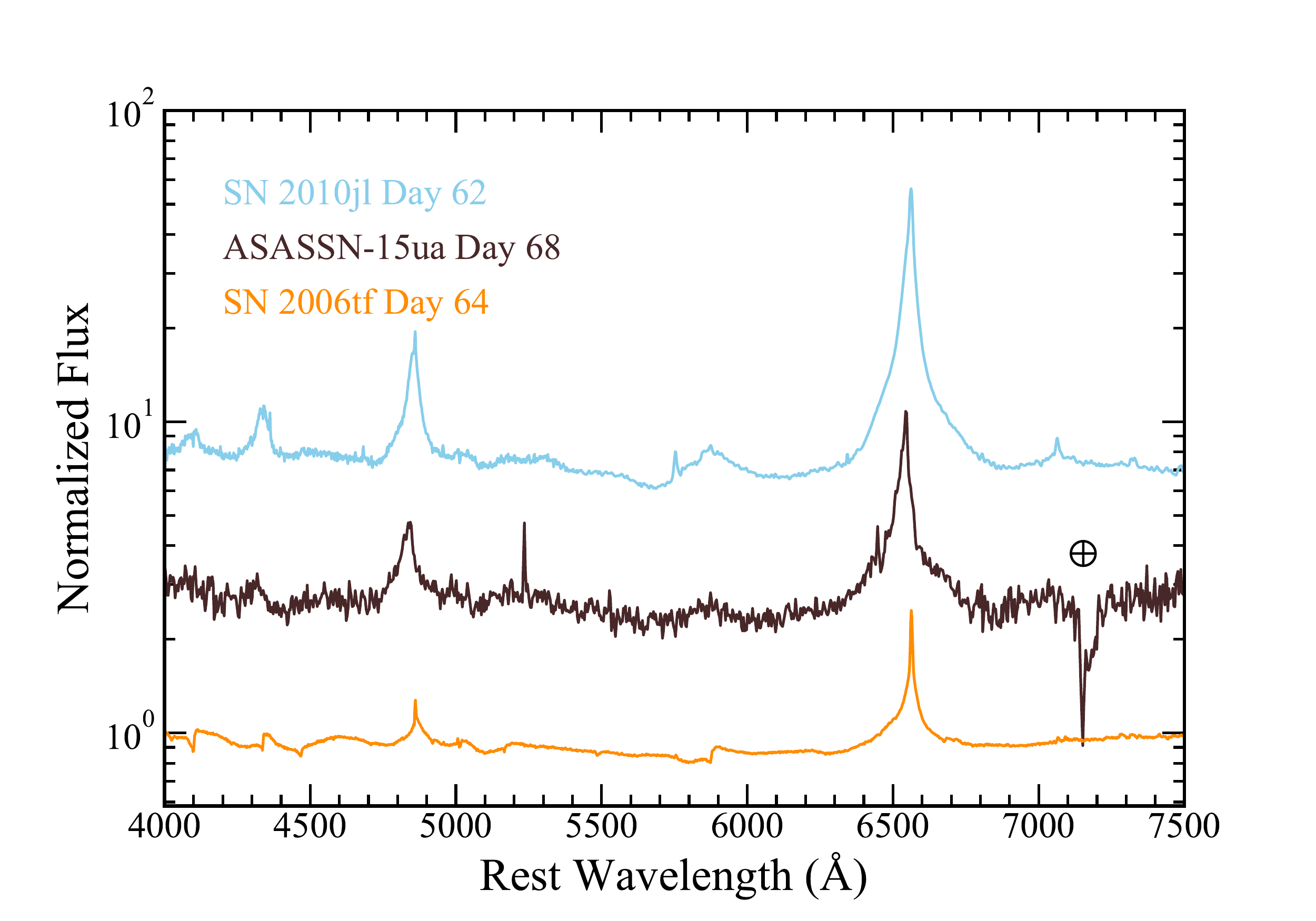}
    \caption{Spectra of ASASSN-15ua, SN2010jl \citep{smith11_10jl}, and SN2006tf \citep{smith2008_06tf} at a similar epoch (68, 62, and 64 days, respectively). All spectra have been corrected for extinction, using $E(B-V)$=0.0247 mag for AASSN-15ua (see above), 0.027 mag for SN 2006tf \citep{smith2008_06tf}, and 0.024 mag for SN 2010jl \citep{jencson2016_10jl}.}
    \label{fig:compare full spec}
\end{figure}

\begin{figure}
    \centering
    \includegraphics[width=\columnwidth]{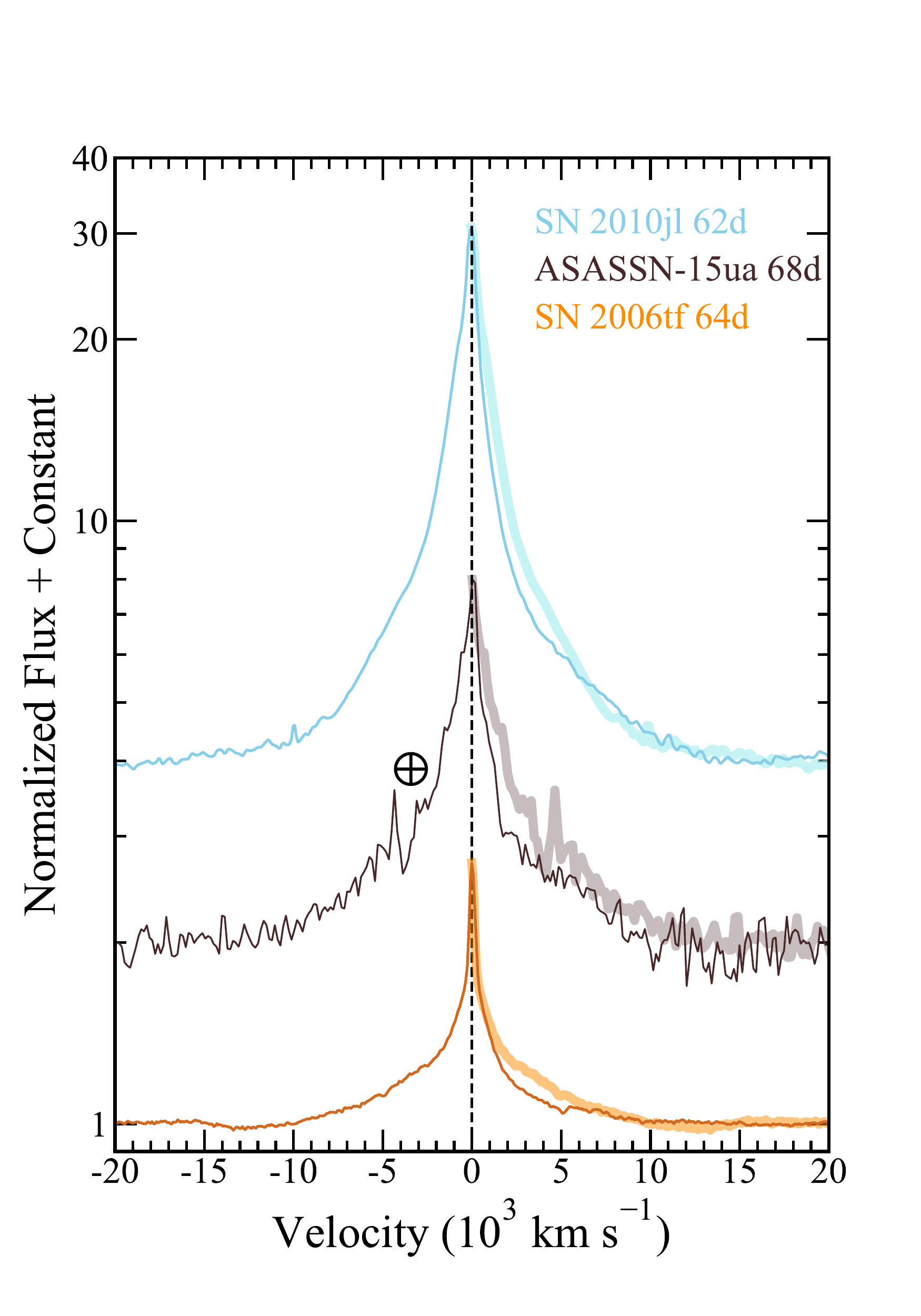}
    \caption{Spectra of ASASSN-15ua, SN2010jl, and SN2006tf at similar epochs (68, 62, and 64 days, respectively) enlarged around H$\alpha$, superimposed and shown with their blue-shifted side reflected to the red side as in Figure~\ref{fig:uaspecreflec}. The thicker and fainter tracing is the reflected blue wing for each SN.}
    \label{fig:compare ha velocity}
\end{figure}

Figure \ref{fig:compare full spec} compares spectra at similar epochs of SN~2006tf, SN~2010jl, and ASASSN-15ua. At $\sim$60 days, electron scattering no longer dominates the H$\alpha$ line profile for all these three SNe. Each object seems to show a slight asymmetry in H$\alpha$, H$\beta$, and even H$\gamma$ line profiles. Most notably, SN~2010jl and SN~2006tf, as well as ASASSN-15ua, have extremely similar spectra at an epoch when the CSM is optically thin, so the underlying ejecta are observable, despite having dramatically different photometric and spectroscopic evolution.

This is also very apparent in Figure \ref{fig:compare ha velocity}, which compares the line profile of ASASSN-15ua to that of the other two SLSNe~IIn. While the line asymmetry in ASASSN-15ua is mild, it is quite similar to the asymmetry seen in both of the two comparison SLSNe IIn at similar phases. We note that in the case of SN~2010jl, the blueshifted asymmetry became progressively significant at a much more dramatic pace at late times \citep{smith2012_10jl,gall14_10jl_dust}. The blueshifted asymmetry was not as pronounced in SN~2006tf \citep{smith2008_06tf}, even at late times.  For ASASSN-15ua, we do not have comparable late-time spectra, but its overall spectral evolution shows a close resemblance to SN~2006tf.

Because of similarities in photometric and spectroscopic evolution between ASASSN-15ua and SN~2006tf, it is unsurprising that its precursor mass-loss history is more similar to SN~2006tf than SN~2010jl. ASASSN-15ua lost mass faster than SN~2010jl, which spent 30 years losing a total of 3M$_\odot$ \citep{Fransson_2014_SN2010jl}. Instead, it is more like SN~2006tf, which spent 20 years losing a total of $\geq$18M$_\odot$ \citep{smith2008_06tf}. The total mass lost by ASASSN-15ua is intermediate between SN~2010jl and SN~2006tf.

Both progenitors of SN~2006tf and SN~2010jl were proposed to have been LBVs or LBV-like eruptive massive stars, and the strong episodic mass loss of ASASSN-15ua places its progenitor in the same parameter space. Our study of ASASSN-15ua indicates that there is a large, continuous range of mass ejection behaviors in these dying massive stars, as discussed below. The total ejected mass can range from 3-25M$_\odot$, after comparing these three SNe IIn alone.

\D{An alternative progenitor scenario is a Red Supergiant (RSG). In SN~2010jl, the trough of the P Cygni profile was 28 km s$^{-1}$, which would be consistent with an RSG-like wind. It is important to note, however, that this is 28 km s$^{-1}$ CSM moving along the line of sight. Most of the outflow emission components were broader than this value (narrow H$\alpha$ emission showed a width of 120 km s$^{-1}$, \citealt{smith11_10jl,smith2012_10jl}).  Since the outflow along our line of sight is substantially different from the bulk expansion speed indicated by isotropic emission, this requires an asymmetric CSM geometry. An outflow speed of 120 km s$^{-1}$ is much faster than expected for a RSG wind. Since these outbursts can lift several solar masses of material off of the star, any RSG progenitor of a SN IIn must be at the extreme end of the ZAMS mass range (see \citealt{smith_2009_betelgeuse_VYCanisMajoris}). }

Assuming a CSM radius of a few $\times 10^{15}$cm, the densities (and mass-loss event timescales) of SN~2006tf, SN~2010jl, and ASASSN-15ua are $0.5-9\times10^{-10}$g cm$^{-3}$ (20 yr), $0.4\times10^{-10}$g cm$^{-3}$ (30 yr), and $1-3\times10^{-10}$g cm$^{-3}$ (12 yr). For both SN~2006tf and ASASSN-15ua, the wind density increased closer to the time of explosion. These values are calculated assuming an isotropic, spherical wind. However, the circumstellar environment observed around LBVs in the MW, LMC, and SMC are often bipolar (\citealt{Smith_2003_etacar_stellarwind}, see also \citealt{smith2020_17hcc}), so an isotropic density measurement is likely not a fair metric of the diversity in mass loss properties among these three objects. 

As discussed by previous studies (for reviews, see \citealt{smith14,smith17}), there are a few potential mechanisms to power these pre-SN mass loss events. 

The Pulsational-Pair Instability (PPI) could drive extreme mass loss, and this mechanism requires a progenitor mass of 100-150M$_\odot$ \citep{woosley07}. The timescale between the last pulse and explosion is highly sensitive to the He core mass with a range of $t=10^{-3}-10^4$yr \citep{woosley_2017_PPISN,leung_2019_PPISN}. The mass loss inferred for ASASSN-15ua is consistent with this range, thus is consistent with PPI. The progenitor of SN~2010jl was estimated to be at least 30M$_\odot$ based on the color of its nearby stellar population. A potential difficulty for explaining ASASSN-15ua with this mechanism is that the high velocities seen in the wings of H$\alpha$ imply the presence of fast SN ejecta, while models of pulsational-pair events typically predict slower bulk expansion speeds around 2,000 km s$^{-1}$ \citep{ws22}. 

Energy deposited into the stellar envelope via wave-driving \citep{sq14,fuller17,wu_fuller_2021_wave_driving} is consistent with the CSM kinetic energy inferred here for ASASSN-15ua. However, the duration of extreme mass loss exceeds the $\sim$1 year duration of Ne and O burning (and far exceeds the $\sim$days of Si burning) that fuels wave-driven mass loss.

The stellar envelope could potentially also become unbound due to a pulsation driven by convective turbulence or explosive burning \citep{sa14}. Energetic pulsations could unbind the remaining H envelope on their own, or sub-binding energy pulsations could be assisted by the progenitor's near-Eddington luminosity to remove the envelope. Lower-energy pulsations or pre-SN swelling of the star's envelope could kick-start mass transfer if the progenitor evolved in close separation to a companion.

In fact, binary interaction is likely to influence or to be responsible for these mass loss events \citep{chevalier_binaryint_2012,smith14,sa14}. Many (2/3 to 3/4 of the population) massive stars in binary systems will interact over the course of their lifetime \citep{sana_2012_binaries_massive_stars, moe_2017_period_mass_ratio_binary}, whereas a small subset of these would need to be experiencing that interaction shortly before explosion in order to produce the nearby dense CSM required for SNe~IIn \citep{sa14}. RLOF and common envelope phases are poorly understood, and the luminosity of progenitors of SNe IIn may approach the Eddington limit, so binary interaction or stellar envelope instabilities may be greatly emphasized.
ASASSN-15ua sheds light on the continuum that exists between these behaviors, and suggests that whatever mechanism powers the pre-SN mass loss in SLSNe IIn should accommodate a wide and continuous range of mass, energy, and pre-SN timescales.


\section{Conclusion}
Our analysis of ASASSN-15ua provides critical insight to the underlying mechanisms driving eruptive mass loss in extremely massive stars and probes the physics of interacting SLSNe. Enumerated below is a summary of this study:

\begin{itemize}
    \item[1.] The superluminous SN IIn, ASASSN-15ua, radiated at least 3$\times10^{50}$ erg. This is an underestimate since the rise to peak and a BC were not accounted for. This value approaches the canonical energy of a typical core-collapse SN, $\sim10^{51}$ erg, suggesting the mechanism behind its extreme luminosity is shock-CSM interaction. 
    
    \item[2.] The progenitor of ASASSN-15ua suffered extreme mass loss for at least 12 years prior to the SN with a steady rate of 0.4 M$_\odot$ yr$^{-1}$, and this rate increased to 0.7 M$_\odot$ yr$^{-1}$ at about 4.5 years prior to SN. The total mass of this CSM is at least 6 M$_\odot$,  imparted with  6 $\times 10^{47}$ erg of kinetic energy. This mass loss was due to an eruptive mechanism, and not a normal line-driven stellar wind.
    
    \item[3.] It is likely that the progenitor of ASASSN-15ua was a Luminous Blue Variable or some similar type of unstable supergiant that cataclysmically lost its H envelope mass, perhaps via binary interaction or pulsational pair instability.
    
    \item[4.] ASASSN-15ua shows the classic dust formation evolution of SNe IIn, where dust began to form in the post-shock gas at most 48 days after discovery and began to block light from the receding portions of the SN, thereby suppressing the redshifted wings of emission lines. This provides additional evidence that radiative cooling in this thin, clumpy region is highly efficient, enabling dust to form.
    
    \item[5.] ASASSN-15ua is an intermediate case between SN~2006tf and SN~2010jl in a number of its measurable properties. ASASSN-15ua's spectral evolution and photometric evolution resemble that of SN~2006tf more closely, while its mass-loss history and continuum evolution are more comparable to SN~2010jl. A similar mechanism may power these objects, being able to produce a range of mass, energy, and timecales in pre-SN mass loss.
\end{itemize}

\section*{Acknowledgements}
We thank an anonymous referee for helpful comments that improved this manuscript. We thank K. Weil and B. Subrayan for helpful guidance and discussion as well as B. Lewis for incredible advice in the development of this work. D.~M.\ acknowledges NSF support from grants PHY-1914448, PHY- 2209451, AST-2037297, and AST-2206532. We wish to extend a special thanks to those of Tohono O'odham ancestry on whose sacred mountain we are privileged to be guests.

Observations using Steward Observatory facilities were obtained as part of the observing program AZTEC: Arizona Transient Exploration and Characterization, which received support from NSF grant AST-1515559. This work was supported through a NASA grant awarded to the Arizona/NASA Space Grant Consortium. Support was also provided by the National Aeronautics and Space Administration (NASA) through HST grant AR-14316 from the Space Telescope Science Institute, operated by AURA, Inc., under NASA contract NAS5-26555. Some data reported here were obtained at the MMT Observatory, a joint facility of the University of Arizona and the Smithsonian Institution.

This work is supported by the international Gemini Observatory, a program of NSF's NOIRLab, which is managed by the Association of Universities for Research in Astronomy (AURA) under a cooperative agreement with the National Science Foundation, on behalf of the Gemini partnership of Argentina, Brazil, Canada, Chile, the Republic of Korea, and the United States of America.

C.D.K. is supported in part by a CIERA postdoctoral fellowship.

Facilities: MMT (Bluechannel), SO: Bok (B\&C), SO: Super-LOTIS.

\section*{Data Availabilty}
The data underlying this article will be shared on reasonable request
to the corresponding author.

\bibliographystyle{mnras}
\bibliography{references}
\label{lastpage}

\end{document}